\documentclass[preprint,12pt]{elsarticle}

\usepackage{amssymb}
\usepackage{amsthm}
\usepackage{lscape}
\usepackage{listings}
\usepackage{graphicx}
\usepackage{multirow}
\usepackage{hyperref}
\hypersetup{
  colorlinks=true,
  linkcolor=blue,
  urlcolor=blue,
  citecolor=black,
}
\usepackage{url}

\journal{arXiv}

\begin{document}

\begin{frontmatter}

\title{Empower Healthcare through a Self-Sovereign Identity Infrastructure for Secure Electronic Health Data Access}

\author[umu]{Antonio L\'opez Mart\'inez\corref{corauth}}
\ead{antonio.lopez41@um.es}
\cortext[corauth]{Corresponding author.}

\author[tsp]{Montassar Naghmouchi}
\ead{montassar-bellah\_naghmouchi@telecom-sudparis.eu}

\author[tsp]{Maryline Laurent}
\ead{maryline.laurent@telecom-sudparis.eu}

\author[tsp]{Joaquin Garcia-Alfaro}
\ead{joaquin.garcia\_alfaro@telecom-sudparis.eu}

\author[umu]{Manuel Gil P\'erez}
\ead{mgilperez@um.es}

\author[umu]{Antonio Ruiz Mart\'inez}
\ead{arm@um.es}

\author[umu]{Pantaleone Nespoli}
\ead{pantaleone.nespoli@um.es}

\affiliation[umu]{
    organization={Department of Information and Communications Engineering, University of Murcia},
    city={Murcia},
    postcode={30100}, 
    country={Spain}
}

\affiliation[tsp]{
    organization={SAMOVAR, Télécom SudParis, Institut Polytechnique de Paris},
    city={Palaiseau},
    postcode={91120}, 
    country={France}
}


\begin{abstract}

Health data is one of the most sensitive data for people, which attracts the attention of malicious activities. We propose an open-source health data management framework, that follows a patient-centric approach. The proposed framework implements the Self-Sovereign Identity paradigm with innovative technologies such as Decentralized Identifiers and Verifiable Credentials. The framework uses Blockchain technology to provide immutability, verifiable data registry, and auditability, as well as an agent-based model to provide protection and privacy for the patient data. We also define different use cases regarding the daily patient-practitioner-laboratory interactions and specific functions to cover patient data loss, data access revocation, and emergency cases where patients are unable to give consent and access to their data. To address this design, a proof of concept is created with an interaction between patient and doctor. The most feasible technologies are selected and the created design is validated. We discuss the differences and novelties of this framework, which includes the patient-centric approach also for data storage, the designed recovery and emergency plan, the defined backup procedure, and the selected blockchain platform.
    
\end{abstract}


\end{frontmatter}



\section{Introduction}
\label{sec:Introduction}
Healthcare  is a key pillar of modern society. It plays a critical role in preventing disease, prolonging life, and promoting health through the organized efforts and informed choices of society, organizations, the public and private sectors, communities, and individuals \cite{deloitte2024}. Within this essential sector, the clinical domain plays a central role, making the management of sensitive patient data indispensable for effectively identifying diagnoses, and critical for treatment and individual health management. The quality of clinical care has a direct impact on patient outcomes and general health. The information managed in this domain, ranging from medical histories to diagnostic results, is not just critical to patient care, but is also invaluable for advancing medical research and shaping public health policy. The European Union proposed the European Health Data Space (EHDS) regulation \cite{eu2022} in recognizing of the importance of efficient data management and sharing. This regulation aims to standardize and enhance data sharing across member states, thereby promoting better clinical and healthcare outcomes by enforcing strong data protection measures and facilitating smoother cross-border healthcare services. This initiative also reinforces the relevance of health data to patient care in the healthcare sector.

Cyberattacks and security issues are some of the most common concerns and threats in the clinical domain. For example, Enzo Biochem, a US biotechnology company, suffered from a ransomware (i.e., encrypting and blocking systems and operations) attack that exposed the clinical test information of 2.5 million patients \cite{EnzoRansomware}. This company detected that attackers exfiltrated this sensitive information, which included 600,000 social security numbers. In this case, the attack vector started with a ransomware attack, but ended with a data breach attack. These two attacks are the most prevalent in the healthcare sector and, subsequently, in the clinical sector. The report that Check Point addressed \cite{CheckPointReport} commented on these attacks, along with malware, distributed denial of service, and phishing. This report also indicated a 60\% increase in these incidents in 2022 compared to the previous year. In terms of data breach costs, IBM released a report indicating since 2020, healthcare data breach costs have increased 53.3\% \cite{IBMReport}. All of these threats and reports highlight the fragility of traditional personal data and identity management systems and the urgent need for more secure and resilient solutions \cite{LopezSurvey2023}. 

In this scenario, Self-Sovereign Identity (SSI) emerges as a groundbreaking solution that potentially redefines data security in the clinical environment. SSI shifts control of digital identities and personal data from institutions to individuals \cite{Soltani2021}. This change has many clinical benefits. In particular, patients can decide who can access their data and for what purposes, significantly enhancing privacy. The decentralized nature of SSI reduces the risks of traditional centralized systems because there is no single point of failure. With this decentralized approach, SSI can be described as an efficient paradigm for protecting and securing clinical assets and data from the threats outlined above \cite{Belchior2020}. In addition, traditional clinical environments have increased interoperability issues due to the variety of machines and devices used in them. In this context, SSI can act as a unifying layer, enabling different systems to interact more seamlessly while maintaining data security and privacy \cite{Yildiz2023}.

One of the primary causes of data exfiltration and data breaches is the centralization of healthcare data. To address this issue, SSI relies on a decentralized infrastructure to implement principles such as ownership, control and autonomy, which are easier to achieve on decentralized infrastructures. For such purposes, blockchain brings several advantages. In particular, this technology serves as a trustless, decentralized public-key infrastructure \cite{BlockchainMotivation}. Blockchain technically manages the identifiers called Decentralized Identifiers (DIDs), as an independent indexing system, from their creation to their revocation. The DIDs are autonomously controlled by the users.  They are the ones who decide about their creation and revocation, not a centralized authority \cite{Tcholakian2023}. DIDs also establish secure peer-to-peer communication channels, providing SSI with a powerful standard. Thanks to DID technology, Verifiable Credentials (VCs) are created to provide digital options for physical credentials. VCs are credentials that contain claims (i.e., pieces of information) about the subject of the VC and can be verified by other entities thanks to cryptographic proofs \cite{VCstandard}. The DID is used here to identify the issuer of the VC. Finally, blockchain also implements Smart Contracts (SCs), which are pieces of code that are automatically executed when certain conditions are satisfied. SCs are executed in a decentralized manner, replicated across blockchain nodes and executed by each node independently, resulting in a more trusted execution. SCs can improve different pillars of the clinical domain, like patient data management and interoperability \cite{Houtan2020}.

To address the problems and limitations identified and the innovative technologies presented, this paper proposes a \textit{novel definition of an open-source SSI framework to protect and secure access to Electronic Health Records (EHRs)}. Our solution uses blockchain, DID, and VC technologies for identification and authentication of users (patients, doctors, laboratory practitioners, etc.), access control, and health data management and sharing. One of the relevant features of this framework is its patient-centric approach, which gives patients full ownership of their health data and stores it on their personal devices, which ensures better privacy and compliance with data protection legislations \cite{eu2022, DataAct}. Moreover, this approach motivates the inclusion of four specific use cases, i.e., i) patient data recovery, ii) data access revocation, iii) verifiable data revocation, and iv) emergency cases. They are needed to cover various challenges such as the loss of personal patient devices, inappropriate use of health data by practitioners, and situations where the patient is unconscious and unable to provide real-time consent for data access. Essentially, our work advances the literature by creating a secure open-source SSI solution that is available for future researchers and developers \cite{GitHubRepository}, by authenticating and authorizing users, and by providing all the required functions for the whole loop of patient data. Moreover, we provide \textit{an implementation of both the blockchain infrastructure and the application servers}, along with a \textit{mobile application} that allows the actors of our use case to create their identifiers, authenticate themselves among themselves and exchange EHRs modeled as VC credentials in a SSI fashion that requires no central authorities and provides a decentralized platform for multiple healthcare actors. We also provide additional functions, such as a recovery protocol for recovering wallet identifiers based on social recovery with the participation of a trusted authority.

The paper is organized as follows. Section \ref{background} provides some preliminary  background. Section \ref{RelatedWork} surveys related work. Section \ref{FrameworkDesign} presents the design of our framework, from the definition of the clinical use case to the framework's design and characteristics. Section \ref{Implementation} shows the implementation of the framework with the technologies selection and its specifications, complemented with a Proof of Concept performed to validate the proposed framework. Performance results are presented in Section \ref{sec:tests}. Section \ref{Discussion} discusses the design presented and the novelties regarding the literature. Section \ref{Conclusion} concludes the paper and outlines perspectives for future work.


\section{Background}
\label{background}
Specific concepts in this paper require a technical background to facilitate the understanding of our proposal. To that end, we present a background of key concepts that will appear throughout the article.

\subsection{Blockchain and Self-Sovereign Identity (SSI)}
Blockchain technology, introduced by Satoshi Nakamoto in 2008 for Bitcoin \cite{nakamoto2008}, is a decentralized digital ledger that records transactions across a network of computers, ensuring security, transparency, and immutability. Each participant in the blockchain network maintains a copy of the ledger, and new blocks of transactions are added through consensus mechanisms like Proof of Work (PoW) or Proof of Stake (PoS). This decentralized nature eliminates the need for a central authority, making blockchain ideal for secure, tamper-proof transactions and data management between multiple stakeholders. Although blockchains provide a trustless infrastructure that can avoid the problems of centralized technologies and provide a shared network, old generations of blockchains --- especially public ones --- lack permission control mechanisms and trusted identity management, which are crucial for their adoption into different domains such as e-health or industrial solutions. Self-Sovereign Identity (SSI), based on newer consortium and private blockchain networks that enable trusted identity management, can ease the adoption of blockchain technology by solving the privacy and performance challenges \cite{Alexandra-Hoss}.

SSI leverages blockchain technology to give individuals full control over their digital identities. Unlike traditional identity management systems that revolve around central authorities, SSI allows users to create and manage their own identifiers, such as Decentralized Identifiers (DIDs). These identifiers are associated with cryptographic keys, ensuring secure and verifiable interactions through the exchange of Verifiable Credentials (VC). By integrating permissioned blockchain with SSI, our framework ensures the integrity, security, and transparency of digital identities and credentials, fostering trust and empowering users to control their personal data (cf. Section \ref{FrameworkDesign}).

\subsection{Decentralized Identifiers (DIDs)}
\label{DIDs}
The W3C DID standard \cite{DIDstandard}, is a type of identifier that enables decentralized, verifiable and self-sovereign digital identities. DIDs are under the full control of their owner, independent from any centralized registry, identity provider, or certificate authority. DIDs are used in conjunction with DID Documents, which contain the public key needed to verify the control over the DID and to verify the signatures on documents (like VC) signed by the DID owner, as well as service endpoints for interacting with the DID owner. In addition, a public DID is discoverable and resolvable, creating a mechanism to reach the DID owner and establishing communication.

There are two types of DIDs: pairwise DIDs and anywise DIDs. Pairwise DIDs are private unique identifiers created for private interactions between two entities. Each interaction or relationship has its own DID, ensuring privacy (non-correlation) and security. Anywise DIDs are public discoverable DIDs that are used across multiple interactions or relationships. Unlike pairwise DIDs, anywise DIDs are not unique to a single relationship and can be reused in various contexts. Anywise DIDs are suitable for public interactions and ideal for public entities like issuers and verifiers, whereas pairwise DIDs are suitable for normal users and private interactions. We envision using both types of DIDs for different purposes (cf. Section \ref{FrameworkDesign}). 

\subsection{Verifiable Credentials and Verifiable Presentations (VCs and VPs)}
VCs and VPs are another W3C standard \cite{VCstandard}. VCs are certified claims (attributes) made by an issuer about a subject's identity and are a fundamental part of decentralized identity systems. These credentials are verifiable in a digital manner, meaning that the entity receiving the credential can be confident in its authenticity and integrity. The key feature of VCs is that they are tamper-proof and can be cryptographically verified. The issuer signs the credential with their private key, and the signature can be checked using the issuer's public key, which is found in their DID Document. This process ensures that the credential is genuine and has not been altered since it was issued.

Verifiable Presentations (VPs) are collections of one or more attributes found in different VCs. VPs are made by the holder of a VC and are presented to a verifier in order to fulfill a request or to demonstrate a qualification, capability, or authority \cite{VPstandard}. If the VC scheme allows it, VPs can implement selective disclosure, where the holder only shares the relevant attributes to the entity requesting them instead of sharing the whole VC. They can also implement Zero-Knowledge Proofs (ZKP) where the holder demonstrates that they have a certain signed attribute without divulgating the attribute itself, or the VC, or even the signature of the issuer, but rather presenting a predicate about the attribute (i.e., Age over 18) and a proof of possession of a VC or signature.
The patient will leverage VPs in our framework to share the minimum possible amount of personal information requested by practitioners and laboratories that allow them to provide health-care while respecting the patient's privacy to a maximum.


\section{Related Work}
\label{RelatedWork}
In this section, we provide a literature review. The purpose is to find specific works for presenting a SSI framework based on blockchain and decentralized technologies in a healthcare/clinical environment. This review focuses on searching the most recent articles and identifying the main concepts and features followed. Table \ref{tab:relatedwork} shows the summary of the works examined, comparing them through different criteria to know if they provide: the blockchain platform used; the data storage approach (patient-centric, centralized, decentralized, etc.) followed; the implementation of Selective Disclosure feature; an emergency plan when patients are unconscious, and they cannot consent to have their health data accessed; the protection of health data using encryption; the implementation of SSI paradigm; the creation of a backup solution to allow patient data recovery in the event of a storage component failure; and, finally, the development of data access revocation to give the patient the ability to revoke access to their data. These functions are important for us because a fully decentralized SSI solution should implement all of them.

Saidi \textit{et al.} \cite{Saidi2022} developed a privacy-preserving decentralized access control scheme based on blockchain and SSI to manage access control and resolve issues of emergencies. The architecture presented comprises three different layers: the Internet of Medical Things (IoMT) devices layer, the user layer, and the Fog and Cloud (F2C) computing layer. They identified an adversary model for their solution, which included a replay attack, spoofing attack, and credential-stuffing attack, as well as security requirements for the designing phase. The authors defined two different access control mechanisms to health data: Role-based Decentralized Access Control (RDAC) and Attributes-based Decentralized Access Control (ADAC). The first one allowed the creation of roles to manage data access, and the second one applied to emergency cases.

\begin{table}[!t]
    \centering
    \caption{Summary of the state-of-the-art studies revised}
    \resizebox{\textwidth}{!}{%
    \begin{tabular}{|*{9}{l|}}
    \hline
    \textbf{Ref./Year} & \textbf{Platform} & \textbf{Data Storage} & \textbf{Selective} & \textbf{Emergency} & \textbf{Encryption} & \textbf{SSI} & \textbf{Backup} & \textbf{Revocation} \\ 
     & &  & \textbf{disclorure} & \textbf{plan} & &  &  &  \\ \hline
    (Bai et al., 2022) \cite{Bai2022} & Hyperledger Indy & Decentralized & -- & -- & \checkmark & \checkmark & -- & \checkmark \\ \hline
    (Harrell et al., 2022) \cite{Harrell2022} & Hyperledger Indy/Aries & Patient & \checkmark & -- & -- & \checkmark & -- & \checkmark \\ \hline
    (Saidi et al., 2022) \cite{Saidi2022} & Hyperledger Indy & Centralized & -- & \checkmark & \checkmark & \checkmark & -- & -- \\ \hline
    (George-Chacko, 2023) \cite{George2023} & Hyperledger Indy/Aries & Centralized & \checkmark & -- & \checkmark & \checkmark & \checkmark & -- \\ \hline
    (Patil, 2023) \cite{Patil2023} & Not specified & Centralized & -- & -- & \checkmark & -- & -- & -- \\ \hline
    (Shuaib et al., 2023) \cite{Shuaib2023} & -- & -- & -- & -- & -- & \checkmark & -- & -- \\ \hline
    (Tcholakian et al., 2023) \cite{Tcholakian2023} & Hyperledger Fabric & Decentralized & -- & \checkmark & \checkmark & \checkmark & -- & \checkmark \\ \hline
    (Yi et al., 2023) \cite{Yi2023} & Hyperledger Fabric & Decentralized & -- & -- & -- & -- & -- & -- \\ \hline
    (Zhuang et al., 2023) \cite{Zhuang2023} & Quorum & Centralized & -- & -- & \checkmark & \checkmark & -- & -- \\ \hline
    (Chintapalli et al., 2024) \cite{Chintapalli2024} & -- & Decentralized & -- & -- & \checkmark & \checkmark & -- & -- \\ \hline
    (Ling-Butakov, 2024) \cite{Ling2024} & -- & Decentralized & -- & -- & -- & \checkmark & -- & -- \\ \hline
    \end{tabular}%
    }
    \label{tab:relatedwork}
\end{table}

Bai \textit{et al.} \cite{Bai2022} implemented a SSI solution for a smart healthcare system. The authors defined the registration and authentication of the different stakeholders (patients, labs, IoMT devices, doctors, etc.). The idea behind Bai \textit{et al.}'s work is to provide a real-time data sharing architecture since they did not present any storage technology to maintain the data, only local storage for the information collected from the IoMT devices, doctors, labs, etc. On the other hand, Harrell \textit{et al.} \cite{Harrell2022} proposed a patient-centric SSI solution. The peculiarity of this solution was the patient data management. The authors defined the data storage in the patient wallets, allowing people to have full control over their data. This evolves one step further the SSI concept. Besides, these authors interestingly conducted the revocation aspect. For instance, when patients provide data to a doctor, they can not remove the data from the doctor anymore, but when they revoke the access, the blockchain stores their action, and the data is no longer verifiable in future shares. However, the authors did not cover important aspects like encryption, emergencies when the patient is unconscious, and backup procedures to avoid data loss.

George and Chacko \cite{George2023} designed a health passport using DIDs and VCs. The authors used DIDs and VCs for the secure creation, sharing, verification, and revocation of generic health VCs. They integrated the solution with a trustable Personal Health Record (PHR) system, collecting the patient data from it without a manual interaction of the user. The authors supported selective disclosure (share only the data requested) and ZKP (cryptographic method to demonstrate a prover has a value without disclosure of any information about the value) in the verification process to enhance user privacy. To provide trust, they had a governance authority agent in charge of issuing the VCs for patients/healthcare organizations and maintaining a list of trusted DIDs of all healthcare organizations. They stored the DIDs, public keys, service endpoints, and proofs in the public ledger to support the validity of the VCs. For data sharing, they used MediTrans, a cloud-based PHR system, which allows secure data sharing and integration with Electronic Medical Record (EMR) systems (presented in Section \ref{sistem-architecture}) thanks to the Fast Healthcare Interoperability Resources (FHIR) protocol, a widely used protocol to exchange health data. This solution maintains patient data in MediTrans \cite{George2022} and EMR systems. When a health record is requested, a VC with the record is created to be shared with the patient.

Zhuang \textit{et al.} \cite{Zhuang2023} incorporated the Non-Fungible Token (NFT) technology to implement secure health data exchange. NFTs are non-fungible tokens that represent digital or physical artifacts storing metadata but not sensitive information about the artifact. The authors used NFTs to represent the patient data, achieving that these tokens were used to manage the access to their data. In this case, the use case developed by Zhuang \textit{et al.} is the data exchange between healthcare providers. The health data are stored in the EMR system, and when the user grants access to a second healthcare provider with its NFT, there is a module of the architecture that encrypts and shares the patient data with another healthcare provider. However, this work did not allow the selective disclosure of health data, only the sharing of all patient data. Shuaib \textit{et al.} \cite{Shuaib2023} presented a short description of SSI concept in healthcare. The authors enumerated the different advantages of this principle: patient control, new links for providers, less risk, compliance, low maintenance, and data available for research.

On the other hand, Tcholakian \textit{et al.} \cite{Tcholakian2023} implemented a SSI solution for consented and content-based access to medical records using blockchain. They proposed an emergency plan using attribute-based encryption, a technique that encrypts data using a set of attributes. They incorporated a hospital emergency server, which had the attributes needed to decrypt patient data when they were unconscious. Besides, they also implemented a data access revocation function, thanks to the attributes used to encrypt the medical record. However, they did not cover any backup solution to prevent data losses. 

Patil \cite{Patil2023} developed a blockchain-based solution to secure patient EHR. This work did not directly implement a SSI solution. \textit{Patil} used different encryption mechanisms: RSA for generating the keys and ABE for patient data's encryption and decryption processes. Some deficiencies of this work are the centralized storage selected and the lack of patient data control. Yi \textit{et al.} \cite{Yi2023} conducted a work to leverage blockchain potential in EHR management. They used the blockchain to store and maintain the medical record, a practice not recommended for this type of sensitive data since all participants in the network can inspect the data.

Chintapalli et al. \cite{Chintapalli2024} proposed a digital signature-based mechanism to improve the security of healthcare data records. They used blockchain and the digital signature mechanism in their design and tested the performance of their solution regarding traditional blockchain solutions. Finally, Ling and Butakov \cite{Ling2024} developed a conceptual trust framework for healthcare metaverse applications. They focused on introducing trust in the metaverse regarding SSI applications. They proposed a basis without implementation and validation.

As depicted in Table \ref{tab:relatedwork}, almost all the works found have been published from 2022, demonstrating a clear recent tendency in SSI solutions in the healthcare context. However, we conclude that there are no previous works incorporating all the functions, criteria, and characteristics presented as columns in Table \ref{tab:relatedwork}. 

The patient-centric approach for data storage, in conjunction with the rest of the functions, could develop a complete SSI solution, defending all the statements established in this paradigm and granting patients real ownership of their health data. In this sense, our framework aims to provide a fully integrated blockchain-based SSI solution, including all the functions revised in the literature, making it open-source, and incorporating laboratories as entities in the platform, allowing the veracity of the clinical results obtained from the patient samples.


\section{Proposed approach for self-sovereign EHR access management}
\label{FrameworkDesign}
\begin{figure*}[ht]
    \centerline{\includegraphics[width=\textwidth]{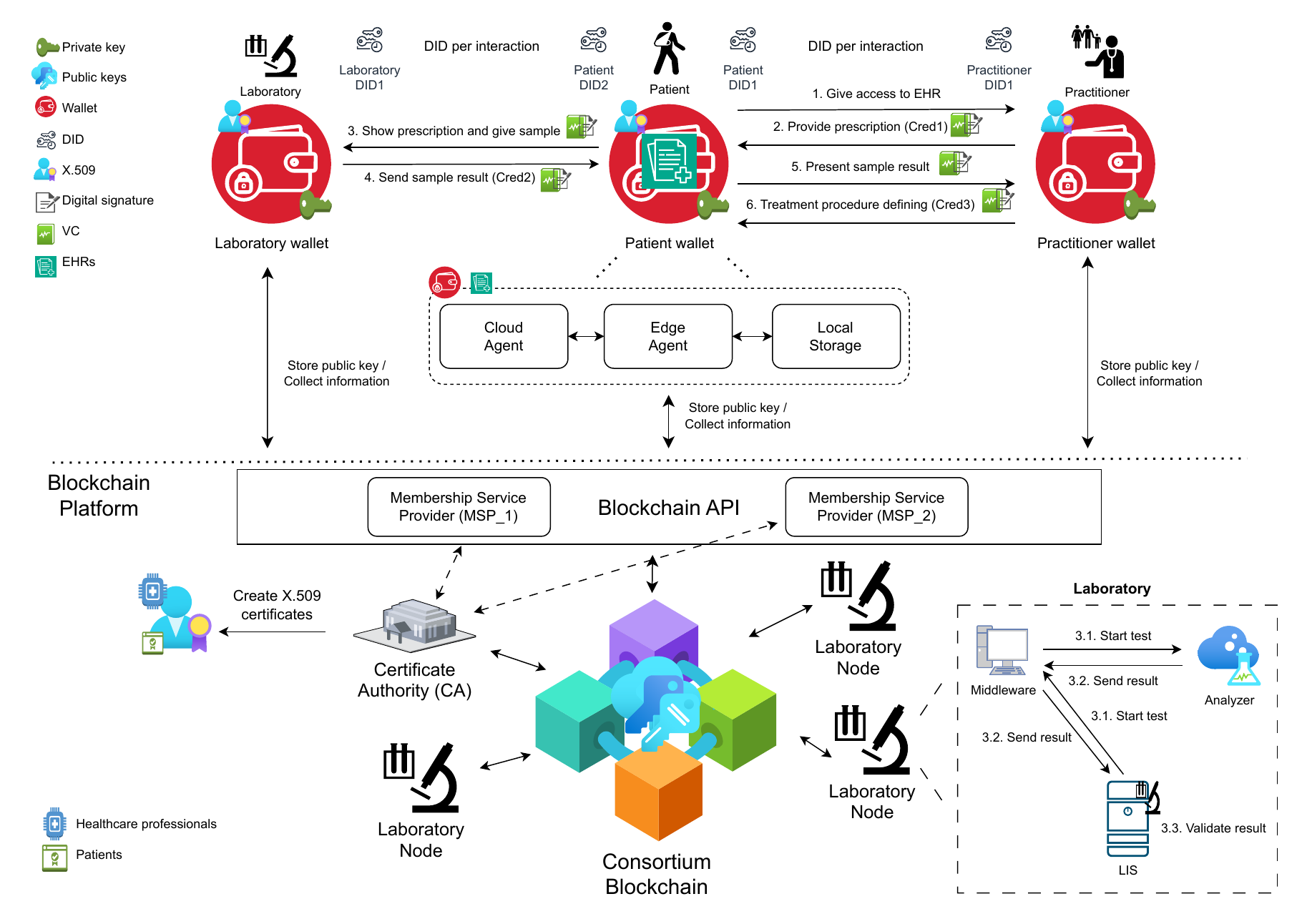}}
    \caption{SSI architecture divided into two parts: (1) user wallet, encompassing all the components directly or indirectly used by the users to interact with the framework, and (2) blockchain platform, to provide authentication, authorization, and verifiable data registry.} 
    \label{fig:Architecture}
\end{figure*}

We propose a Self-Sovereign Identity (SSI) framework to protect and secure access to EHRs in a clinical environment. We envision this proposal as patient-centric since the EHRs are uniquely stored in the patient's wallet, providing full control and management of these data. Figure \ref{fig:Architecture} provides a complete overview of the framework architecture and interactions. Essentially, the architecture is divided into two parts: the user wallet and the blockchain platform.

\subsection{System model}
\label{sistem-architecture}

\subsubsection{User Wallet}
As Figure \ref{fig:Architecture} shows, this framework is designed to be used daily for medical appointments and to analyze patient samples, which occurs when the patients have an illness or request an analysis. In this sense, we identify three main users: the patient, the practitioner, and the laboratory. Firstly, the patient is the user that owns and manages the health data, the practitioner is the healthcare professional who performs a treatment procedure, and the laboratory is represented by the laboratory professionals who extract and analyse the patient samples.

The three users interact with the framework and with each other through a personal wallet application. This wallet implements specific functionality for the different use cases addressed in the framework, which we introduce in Sections \ref{standard-usecase} and \ref{framework-characteristics}.

The user wallet comprises three components: edge agent, cloud agent, and local storage. In the edge agent, we include all the functionality related to creating DIDs, managing public and private keys, signing VCs, and reading blockchain data for necessary verification. This functionality manages with the specifications of the SSI paradigm, giving patients control over all information and processes about their identity and data, the main advantage of this component. The edge agent is hosted in the personal device of the entity. The cloud agent intervenes whenever an edge agent needs to connect to another wallet or the blockchain platform. We can use the cloud agent to implement specific functions, such as the backup and data access revocation of health data. The edge agent will leverage the cloud agent to extend its capabilities exponentially.  as we will present in the following subsections. In this case, the cloud agent is external to the personal device. Finally, we include the local storage module, where we will securely store the patient EHR in the form of VCs, DIDs, and public and private keys (the private key can be stored outside the wallet).

In addition, the main functionality for laboratory and practitioner wallets is to generate VCs that represent prescriptions, diagnoses, and laboratory results and send them to the patient's wallet. The laboratory contains specific components, i.e., analyzers, middlewares, and Laboratory Information Systems (LIS) \cite{Lopez2023}. The analyzers are the laboratory machines that obtain the result from the physical patient sample (blood, urine, etc.). This machine is proprietary and uses its particular language to represent the result obtained. Therefore, when a company installs a new analyzer in a laboratory, it also provides the middleware. Laboratory professionals use the middleware to monitor the analyzer, configure it, and see the results obtained. At a higher level, we find the Laboratory Information System (LIS). LIS aggregates the data collected from the different middleware components and validates all the patient sample results (e.g., correct format, no inconsistent data, etc.). This workflow is shown at the bottom right of Figure \ref{fig:Architecture}. 

This framework eliminates one of the central components of the traditional clinical environment, the Electronic Medical Record (EMR) system. The EMR system is the centralized point of the clinical scenario for collecting patient data. We propose that each patient's data be concentrated on their personal device. The specific laboratory entities have no direct connection with the framework, only to generate the patient sample result. The laboratory wallet triggers the patient sample analysis and collects the final results.

\subsubsection{Blockchain platform}
The next part of the framework is the blockchain platform, which provides permanent tamper-proof transactions, decentralization, and shared governance while maintaining data integrity. In the blockchain platform, we include the blockchain API needed to communicate with the blockchain and store and read data from it. 

For accessing patient data, practitioners and laboratories need to have a relationship of trust with the patient, who gives access to their data for a treatment procedure definition. This trusted relationship is achieved thanks to the blockchain solution since we implement a permissioned blockchain, where only authorized entities can participate \cite{Helliar2020}. To do that, the blockchain solution contains the concept of organization, understood as a group of members belonging to a blockchain node. Considering the framework's users, we could have an organization with patients, practitioners, and laboratories in a physical zone. To enter this organization, we envision a Certificate Authority (CA), which provides a certificate if the user meets the requirements and he/she uses such a certificate for authenticating into the framework.

Secondly, we have the authorization part. For that, the infrastructure provides the Membership Service Provider (MSP) component, which translates the certificate received into a specific role (with certain established permissions) inside the organization. The MSP is in charge of assigning the role/permissions of patients, practitioners, and laboratories regarding the certificate provided by the user. Therefore, the MSP, in conjunction with the CA, acts as the root of trust for the framework.  

To conclude, we find the consortium blockchain. We select a consortium blockchain since it is permissioned, and this scenario requests that only authorized users can join the framework. The decentralized database (ledger) is used for storing different information: i) public keys of framework entities, using the ledger as a registry of trusted entities; ii) records with information about special accesses given to stakeholders in emergency cases; and iii) hashes of offline data that we can maintain their integrity.

Furthermore, the blockchain ledger implements storage channels. These channels are established by a sub-group of blockchain nodes, and only accessible to users authorized by the MSPs, chosen by the nodes that created the channels. For instance, we might have a patient-related channel inside the blockchain ledger where information about the accesses given to practitioners is stored. This information is irrelevant for other entities appearing in the framework, like a laboratory, so we can create specific channels between entities where managing the access to the blockchain and only information relevant to them is stored. These channels also produce other functions, such as assigning SCs to specific channels.

All the abovementioned entities conform to the core part of the proposed framework. However, it could be possible to extend the framework to include other external users easily. For instance, we could include pharmacies and insurance companies, allowing patients to show the prescriptions they received from doctors to get the specific medication and pay for the clinical services they received, while ensuring irrefutable proof to be used later to get coverage from their insurance companies. Every new actor that wants to be added, should have a wallet and a certificate to access the permissioned blockchain and request data to verify the information shown by the patient. This considered function makes our framework easily extensible for other/new use cases.

\subsection{Design processes for the clinical use case}
\label{standard-usecase}
Our framework comprises four main design processes. These phases are: i) creation of a user certificate; ii) authorization of a user and assignment in the correct channel(s); iii) mutual authentication between users; and iv) exchange of information between users. We further explain each of these phases in the following subsections.

\subsubsection{Creation of user certificate}
\begin{figure}[ht]
    \centerline{\includegraphics[width=10cm]{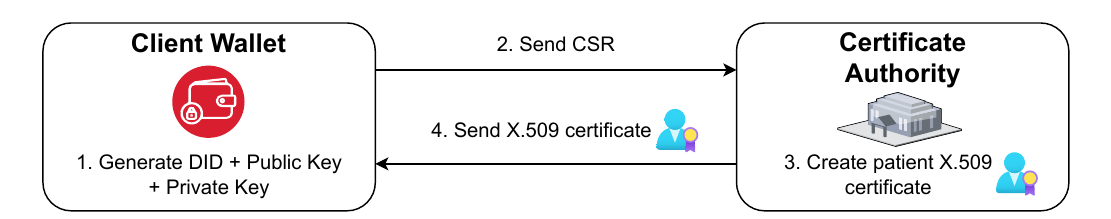}}
    \caption{Creation of a user certificate.} 
    \label{fig:certificate-creation}
\end{figure}

This process, shown in Figure \ref{fig:certificate-creation}, is common to all users (patient, practitioner, and laboratory) with personal wallets. Here, the user identity is authenticated (internal checks and identity proofing are done), and a certificate is provided to give access to the framework. The first step is to generate a cryptographic key pair consisting of private and public keys. The private key is securely stored, either outside or inside the user's wallet, and is never shared. The user uses this key to sign information and requests, to prove their identity. The user also creates an anywise DID. The DID is associated with the previously created key pair. This DID serves as an identifier that can be used to identify and authenticate the DID owner and will be associated with the user's certificate (see step 2). Therefore, this DID and the public key are stored in the blockchain platform to recognize the user in the framework.

The user creates a Certificate Signing Request, known as a CSR (step 2). This request is signed with the user's private key and sent to the CA. At this point, the CA receives the CSR and verifies the user's identity and the authenticity of this request. The CA verifies that the signature provided in the CSR matches the public key associated with the user's DID, found in the DID document. After this verification, the CA issues an X.509 certificate (step 3), which is used to authenticate the user as a legitimate user in subsequent user-ledger interactions in the framework. Also, the CA stores a match between the patient's DID and their real identity to be prepared for an emergency case, as we will present in Section \ref{emergency-case}. Finally, the CA sends the X.509 certificate in response to the user's request (step 4). This certificate is relevant because it associates the user with specific permissions in the framework, channels in the ledger, etc. This process is important because our framework is permissioned, so no one can read/write from the blockchain platform. Therefore, the CA and the MSP become critical to manage the framework's access rights, and permissions.

\subsubsection{User authorization}
\begin{figure}[ht]
    \centerline{\includegraphics[width=10cm]{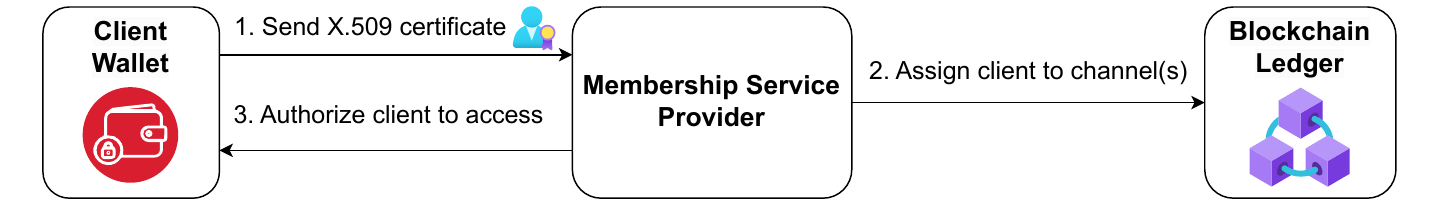}}
    \caption{User authorization.} 
    \label{fig:client-authorization}
\end{figure}

In this phase, the user wallet presents its previously received X.509 certificate to the MSP. This process occurs when the user is logged in to the framework for the first time. Now, the MSP component validates the authenticity of the certificate and checks the permissions (admin user, patient user, practitioner user, etc.) that the certificate owner has within the framework. This verification is performed based on the rules and parameters configured by the MSP. At this point, specific rules can be set, for example, for foreign patients visiting the country. After that, the user is assigned to specific storage channels, another layer of security and privacy. As shown above, these channels provide isolation between different user roles using the framework. At the end, the user receives the acknowledgment and is prepared to interact with the blockchain platform in the correct manner. Figure \ref{fig:client-authorization} visually represents this process. This process allows us to provide a decentralized and scalable authorization process, for example, by replicating MSPs in different locations.

\subsubsection{Mutual authentication}
Mutual authentication involves establishing a mutually authenticated channel between users, such as patients and doctors. First, both patient and doctor wallets generate a pairwise DID, used to protect the communication between them. For example, the patient starts creating a QR code, and the doctor scans such a code to read the pairwise DID created by the patient. The patient then receives the DID from the doctor via the cloud agent, since the doctor receives the information of the patient's cloud agent from the patient's DID. Furthermore, the patient verifies the doctor's identity using the public key associated with the doctor's DID, which is available in the trusted ledger registry (i.e., by creating an authentication challenge to be solved using the associated private key). Now that the patient has authenticated the practitioner, in some cases, the practitioner would also like to authenticate the patient, too, using the patient's certificate and anywise DID (also registered on the blockchain ledger), and in other cases, such authentication is not required for privacy reasons, using only the pairwise DIDs created between them.

\subsubsection{Information sharing}
\label{information-sharing}
\begin{figure}[ht]
    \centerline{\includegraphics[width=11cm]{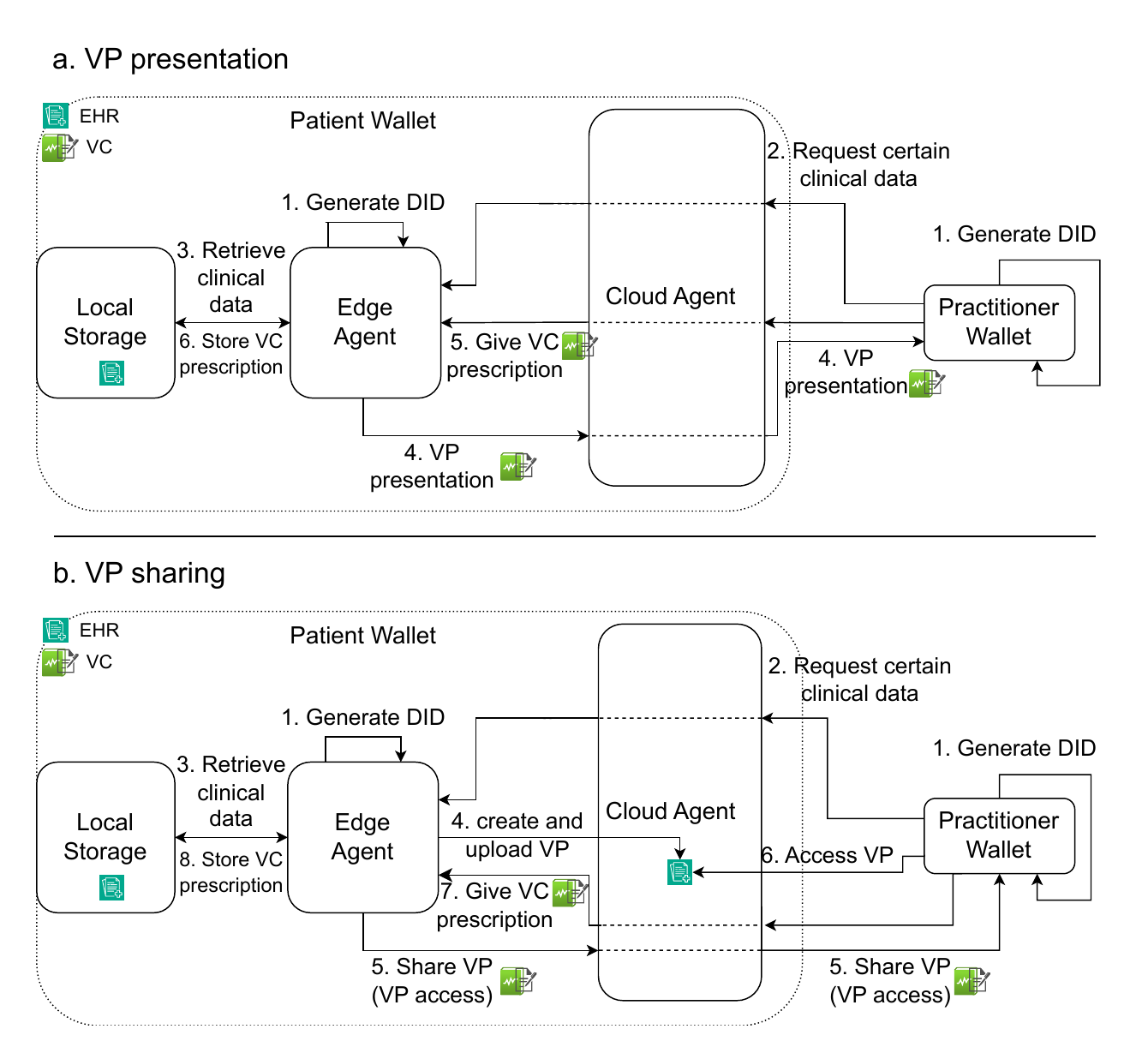}}
    \caption{Information sharing alternatives.} 
    \label{fig:information-sharing}
\end{figure}

Information sharing includes the normal information flow produced in a medical appointment. Figure 4 shows the flows produced for the two alternatives for sharing information between the patient and the practitioner. In the first case (see Figure \ref{fig:information-sharing}a), the practitioner physically reads the data from the patient's device. In this case, the patient generates a Verifiable Presentation (VP), which is a presentation of the data the patient wants to demonstrate to the practitioner (i.e., blood analysis value, previous diagnosis). Such data includes the digital signatures of the entities that created the data (e.g., lab, other practitioners), making it easily verifiable by the doctor through a QR code created by the patient. This case is ideal for quick in-person data sharing where the doctor has no need to keep the data and only a wider idea about the patient's condition.

For the second option (see Figure \ref{fig:information-sharing}b), the patient's wallet creates a VP with selective disclosure of the health data relevant to the doctor and uploads it to the cloud agent. This option allows remote consultations and shares the patient data for a certain period of time. The patient then grants the doctor access to this VP on the cloud. This access can be temporary, expiring after a user-selected time period (similar to location sharing on social media) and can also be revoked by the patient at any time. We will explain the revocation processes created for both alternatives (Sections \ref{vp-revocation} and \ref{vc-revocation}). This second case is ideal for cases where patients perform a remote consultation (i.e., telemedicine) or where doctors need longer time to consult a patient's EHR.

After accessing the patient's health data, the doctor provides a VC containing his diagnosis or the prescription, which can be verified by other entities. To correctly model a clinical standard use case, this prescription requires the patient to extract a sample in a laboratory. The patient's wallet receives the prescription and adds it to the locally stored EHR (two final steps of two alternatives in Figure \ref{fig:information-sharing}). To continue, the patient goes to the lab for the physical sample extraction. Similarly, patient and lab wallets generate a pairwise DID, creating secure communication, and the patient shows the VC (with the prescription) signed by the doctor. The lab checks the authenticity and identity of the patient and doctor. Then, a nurse physically extracts the sample, which is electronically initialized by LIS and middleware components presented in Section \ref{sistem-architecture}, and physically analyzed with an analyzer. With the result obtained, the lab sends the analysis results and data to the patient's wallet as a VC. The lab generates this VC.

The patient presents the VC created by the laboratory with the sample result to the doctor. At this point, the doctor creates a treatment procedure, and the patient receives it. Finally, there could be different actors involved, such as pharmacies and insurance companies, to provide the medication and evaluate the cost of the tests performed on the patient. Thanks to this process, the patient can autonomously manage and share their data, with security and privacy provided by the cloud agent and the blockchain platform.

\subsection{Additional framework functions}
\label{framework-characteristics}
Besides the clinical use cases presented, our framework covers four functions: i) Patient wallet recovery (Section \ref{wallet-recovery}), using the cloud agent shown in Figure \ref{fig:Architecture}; ii) Health data access revocation (Section \ref{vp-revocation}), when the user revokes access to their EHR data; iii) VC revocation (Section \ref{vc-revocation}), to include cases where the issuer wishes to revoke a VC, such as when a practitioner revokes an incorrect prescription; and iv) an emergency case (Section \ref{emergency-case}), when the user can not personally provide access to their data but it is important for the patient's safety to access the information, as opposed to an appointment. We describe these functions through different scenarios in the following subsections.

\subsubsection{Patient wallet recovery}
\label{wallet-recovery}
The patient's personal device plays a crucial role in the framework. If something happens to this device, such as physical damage, loss, or modification of the device, the patient's data may be compromised. Therefore, we envision a backup storage to recover the patient data, and a way to recover the private keys, which will be explained later. The backup created with the patient data is stored in the cloud agent shown in Figure \ref{fig:Architecture}. In this component, health data is kept private, encrypted with a patient's public key, and protected from malicious attempts and actors. Besides, a hash of the backup is stored in the blockchain ledger to ensure the integrity of this information. This hash is calculated and published on blockchain by the patient device before the encrypted backup is sent to the cloud agent, ensuring that the retrieved data is intact and has not been modified at any point while stored on the cloud agent. 

Recovering the private key to decrypt the backup data or other private keys created and stored on the edge agent, which are associated with the DIDs the user creates, is done by implementing a social recovery protocol. This protocol splits the private key and stores chunks in different locations, mainly entrust them to trusted parties (contact family) chosen by the patient \cite{horcruxes, key-rec}. The protocol is easily integrated into our framework since we rely on emergency contacts and the MSP to store key-parts (see subsection \ref{emergency-case}). For simplicity, a key-pair is created when the patient opts for the backup and emergency loop option. The public key is used to encrypt the wallet data and periodically upload it to the cloud agent. The private key is stored securely on the patient's device. It is also split in half; one half is sent to the MSP for safekeeping, along with the emergency contact list; the other half is duplicated and sent to trusted contacts for safekeeping. This social recovery combines an authority (i.e., the MSP) and normal contacts.

When patients want to recover their wallet and data, they will query the MSP to perform the necessary checks before providing them the first half-key. Later, patients contact the rest of their social contacts who hold the other half, which are the same emergency contacts described in the emergency loop below. After receiving the key-parts, patients are able to reconstruct their private key, download and decrypt the data from the cloud agent, which means they end up with an edge agent that has the backed-up credentials, a certificate from their MSP, and their main private keys. However, any connections made with pairwise DIDs must be reestablished with new identifiers since they were not backed up. Nevertheless, this extra step is not a problem, since periodic updates and changes to such connections are always desired in an identity system to improve security and privacy.

\subsubsection{Health data access revocation}
\label{vp-revocation}
Our framework stores health data on the patient's device. Nonetheless, managing access to such information is a challenge. To address such a challenge, we must first explain how we envision health data access in the framework for defining data access revocation. As described in the \textit{Information sharing} process (Section \ref{information-sharing}), the patient can locally show the health data to the practitioner, like the traditional approach where we show practitioner our medical documents, with a difference that they can verify the integrity and authenticity by scanning a QR code that we create when sharing. 

\begin{figure}[ht]
    \centerline{\includegraphics[width=10cm]{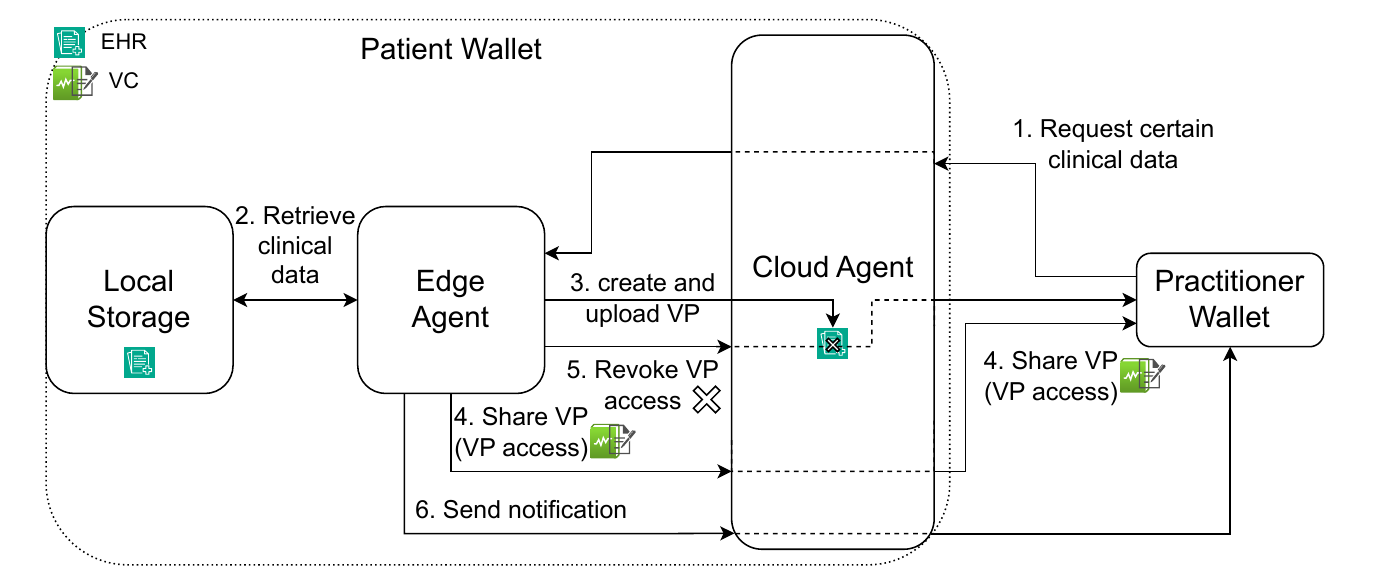}}
    \caption{Data access revocation workflow.} 
    \label{fig:data-access}
\end{figure}

On the other hand, when VP is shared, we provide access to a VP created on the local device and uploaded to the cloud agent. This process is performed entirely by the user application, with complete abstraction from what happens outside the scene. Figure \ref{fig:data-access} presents the revocation workflow. When the patient wallet receives the request for certain health data (\textit{step 1}), it retrieves the data from the local storage (\textit{step 2}), creates and uploads a VP with the relevant data in the cloud agent (\textit{step 3}), and shares VP access with the requester's wallet (\textit{step 4}), in accordance with what was described in Figure \ref{fig:information-sharing}. Finally, the patient's wallet has the ability to revoke access to this VP, by removing it from the cloud agent (\textit{step 5}) and notifying the requester's wallet of the termination of the authorization by the cloud agent (\textit{step 6}). Thanks to this function, patients can manage their data with full control.

\subsubsection{VC revocation}
\label{vc-revocation}
Besides health data access revocation, we may need to revoke a VC. For instance, if a doctor has issued a VC with a wrong prescription or a lab has issued a VC with a patient test results that needs to be repeated. To achieve it, we dive deeper into the concept of verifiable credentials (introduced in Section \ref{background}) and the revocation registry element.

\begin{figure}[ht]
    \centerline{\includegraphics[width=10cm]{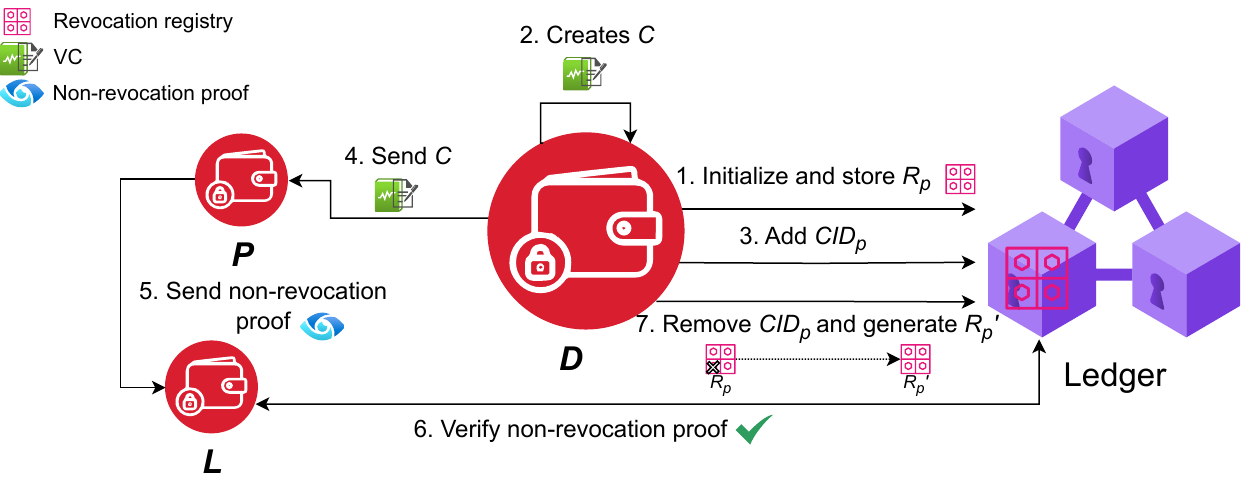}}
    \caption{VC revocation workflow.} 
    \label{fig:vc-revocation}
\end{figure}

First, consider a scenario where a patient \textit{P}, a doctor \textit{D}, and a laboratory \textit{L}, as shown in Figure \ref{fig:vc-revocation}. For this example, we assume that the doctor \textit{D} is the creator of the credential \textit{VC} and is responsible for revoking it. Thus, \textit{D} initializes a revocation registry \textit{$R_P$}, stored in the blockchain ledger, which contains a registry of non-revoked credentials issued by \textit{D}. When \textit{D} issues a \textit{VC} with certain information for \textit{P}, \textit{D} updates \textit{$R_P$} in the ledger, indicating that \textit{VC}, with a unique identifier \textit{$CID_P$}, has been issued and marked as not revoked. If \textit{D} revokes the credential \textit{VC}, \textit{D} has to update \textit{$R_P$}, eliminating \textit{$CID_P$} of \textit{$R_P$} and computing a new state of \textit{$R_P$}, called \textit{$R_P$'}. This fact is due to the design of the revocation registry,i.e., an accumulator. This data structure allows many values to be represented in a single constant-size value, where it is possible to add and remove values and generate proof that a specific value is contained in the accumulator without revealing any other values. \textit{D} also signs this new state \textit{$R_P$'} produced in the registry so that is can be verified by \textit{P} and \textit{L}.

Now, if \textit{P} wants to exchange the VC received from \textit{D} with \textit{L} or other entities of the framework, \textit{P} needs to present a non-revocation proof to \textit{L}. In Section \ref{Implementation}, we will explain the technology used to implement this non-revocation proof. Finally, if \textit{D} revokes \textit{VC}, the other entities can no longer use this VC in a trusted way, ensuring that the VC owner can manage the VC created. This function avoids impersonations or malicious uses of VCs since the VC needs to be active, and the user can easily revoke the VC. 

\subsubsection{Emergency case}
\label{emergency-case}
In an emergency situation, where the patient is unconscious or unable to provide EHR access, the doctor needs access to the patient's data in order to perform a preliminary triage and assess the severity of the situation. However, in the proposed framework, such information is stored locally in the patient's device, so the doctor can not obtain consent to view the health data. To solve this problem, we define a special case where the doctor could access patient data without real-time consent.

When the unconscious patient arrives at the hospital, the caregiver creates a request to the MSP, specifying the identity of the patient\footnote{We assume we can identify the patient through the patient's device or through their available identity documents} and requesting access to patient's data. The MSP receives this request and starts the emergency loop. For a proper use of this loop, three particular prerequisites must be met. First, the patient must have accepted the backup mechanism by having a health data backup encrypted with a public key in the patient's cloud agent. Second, the patient must have provided the MSP with a set of trusted contacts (e.g., trusted doctor, family member, etc.) at the time the X.509 certificate was created or when the backup mechanism was triggered. Finally, the patient wallet must have split the private key used to encrypt the health data backup, sending one part to the MSP and the other part redundantly to the trusted contacts (during the recovery set-up described in Section \ref{wallet-recovery}). The MSP will ping them in a given order until it receives the other part of the key. This way, neither part can decrypt the data without the other part.

With these preliminaries satisfied, the MSP initiates the loop, contacting trusted contacts to obtain the other part of the private key. The wallet may receive a notification requesting the key part of the person's private key who trusted you. Once a person responds, the MSP sends the private key and the information to the doctor to reach the patient cloud agent. At the same time, the MSP creates a record with the information of this emergency loop triggered. This record is stored in a special security channel in the ledger. A record  of the emergency loops triggered with  patient and doctor's information is maintained in this security channel. This data will be critical for audibility and security purposes, as the MSP could improperly initiate a simulated loop to receive the other key part and access patient data.

Once the access is given to the doctor, the doctor will perform a request to receive patient's health data. In this sense, the cloud agent should manage the given accesses and a registry of the information provided to the doctor. For this purpose, the cloud agent can implement \textit{A Posteriori Access Control} capabilities that allow the doctor to receive the information, while monitoring and auditing all actions perfoemd, in order to apply sanctions if policy violations are detected \cite{Galvan2021, Azkia2015}.

However, the patient may not have accepted the backup mechanism. If this occurs, the patient may have shared a contact list with the MSP. In this case, the doctor can receive such a contact person and trust their information. This case is intended for patients who do not trust cloud storage and prefer to keep their health data locally on their devices. Finally, if the patient does not accept the contact mechanism either, the doctor must evaluate the patient's severity and create the treatment procedure without the patient's EHR.

\section{Implementation and specifications of the framework}
\label{Implementation}
\begin{figure}[ht]
    \centering
    \includegraphics[width=8.5cm]{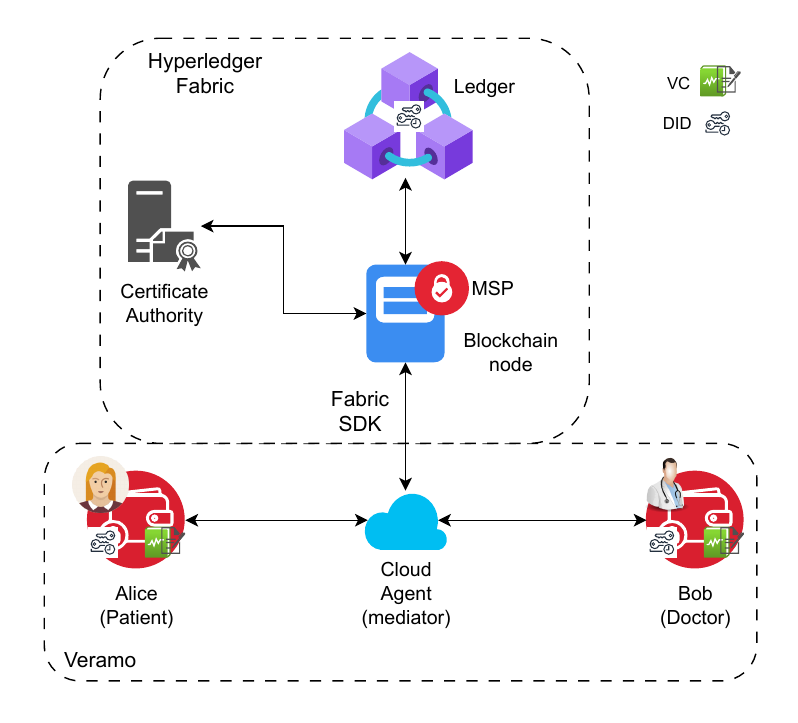}
    \caption{Proof of Concept design.} 
    \label{fig:proof-of-concept}
\end{figure}

The design proposed in Section \ref{FrameworkDesign} encompasses many aspects of the framework. In this section, we present the technologies we chose to deliver a Proof of Concept (PoC). This PoC contains the implementation of key functionalities required for the functions and processes presented above: (i) deploying blockchain platform and Smart Contracts to store and retrieve data; (ii) implementing edge agents; (iii) providing a cloud agent (mediator); (iv) creating mutual authentication and peer-to-peer communication channels between edge agents through cloud agent; and (v) creating, storing, verifying and managing DIDs and VCs. The following sections present specifications of each core component of the framework and the PoC scenario proposed in Figure \ref{fig:proof-of-concept},.

\subsection{Specifications of the core components}
\subsubsection{Blockchain platform}
\label{blockchain-platform}
Starting from the bottom of Figure \ref{fig:Architecture}, it is possible to see the blockchain platform used for the PoC. In this sense, we conducted a complete review, reviewing the literature to obtain the most used open-source blockchain platforms in healthcare \cite{Merlo2023, J2023} and extracting the features to compare them \cite{Farshidi2020, Nanayakkara2021, Capocasale2023}. Table \ref{tab:blockchainplatforms} presents the results of this research. The relevant features extracted from the literature for our analysis are: (i) the purpose of the blockchain platform; (ii) the network type; (iii) the capability of implementing SCs, (iv) membership services (MSP), and (v) storage channels; (vi) privacy; (vii) SSI-focused; (viii) documentation (\checkmark for vague documentation, \checkmark\checkmark for normal documentation, and \checkmark\checkmark\checkmark for excellent documentation); and (ix) performance. These are the most important features given the proposed use case and the SSI-focused framework designed.

\begin{table*}
    \centering
    \caption{Blockchain platforms comparison for the proposed framework.}
    \resizebox{\textwidth}{!}{%
    \begin{tabular}{|*{10}{c|}}
    \hline
    \textbf{Platform} & \textbf{Purpose} & \textbf{Network type} & \textbf{SCs} & \textbf{MSP} & \textbf{Storage channels} & \textbf{Privacy} & \textbf{SSI} & \textbf{Doc.} & \textbf{Performance} \\ \hline
    Bitcoin & Financial & Permissionless & -- & -- & -- & -- & -- & \checkmark\checkmark & Low \\ \hline    
    Hyperledger Fabric & Industrial & Permissioned & \checkmark & \checkmark & \checkmark & \checkmark & \checkmark & \checkmark\checkmark\checkmark & High \\ \hline
    Ethereum & Industrial & Permissioned & \checkmark & -- & -- & \checkmark & \checkmark & \checkmark\checkmark\checkmark & Medium \\ \hline
    Hyperledger Indy & DID & Permissioned & -- & -- & -- & \checkmark & \checkmark & \checkmark\checkmark\checkmark & Medium \\ \hline
    Hyperledger Sawtooth & Industrial & Permissioned & \checkmark & -- & -- & \checkmark & -- & \checkmark\checkmark\checkmark & High \\ \hline
    Quorum & Industrial & Permissioned & -- & -- & -- & -- & \checkmark & \checkmark & Medium \\ \hline
    Corda & Financial & Permissioned & \checkmark & -- & \checkmark & \checkmark & -- & \checkmark\checkmark\checkmark & High \\ \hline
    Hedera & Industrial & Permissionless & \checkmark & -- & -- & -- & -- & \checkmark\checkmark\checkmark & High \\ \hline
    Cosmos & Industrial & Permissionless & \checkmark & -- & \checkmark & \checkmark & -- & \checkmark\checkmark\checkmark & High \\ \hline
    \end{tabular}%
    }
    \label{tab:blockchainplatforms}
\end{table*}

Inspecting Table \ref{tab:blockchainplatforms}, Hyperledger Fabric and Hyperledger Indy are the best options for our work. They are two open-source projects created by Hyperledger Foundation. Fabric \cite{HyperledgerFabric} is a permissioned blockchain platform designed to develop enterprise-grade solutions that require a high degree of privacy, scalability, performance, and flexibility. It offers a modular architecture that enables the integration of consensus algorithms, membership services, and smart contracts to meet the specific needs of a business network. In contrast, Indy \cite{HyperledgerIndy} is a distributed ledger purpose-built for decentralized identity management. It provides tools and libraries for creating and managing digital identities rooted in blockchain technology, enabling users to control and share their personal information securely. Indy is specially focused on SSI use cases. And, as presented in Section \ref{FrameworkDesign}, we envision our design as a SSI solution, but leveraging the capabilities of smart contracts, storage channels, and membership services, among others, to satisfy different processes inside healthcare, not just identity management. Therefore, we select Hyperledger Fabric as the blockchain platform to be used for our framework.

\subsubsection{User Wallet}
The user wallet contains the functionality of the Edge Agent and Local Storage shown in Figure \ref{fig:Architecture}. In this context, we considered two alternatives for the implementation: Hyperledger Aries and Veramo. On the one hand, Aries \cite{HyperledgerAries} is a project focused on creating interoperable implementations of decentralized identity protocols. It provides a toolkit for building secure, scalable, and privacy-oriented identity applications. However, Aries relies heavily on the Hyperledger Indy blockchain, making it difficult to replace Indy with the Fabric blockchain. 

On the other hand, Veramo \cite{Veramo} is a framework that provides developers with the tools to build applications using DIDs and VCs. It provides a flexible, modular architecture that supports various identity protocols and blockchain technologies. This flexibility in the selection of identity protocols and blockchain technologies makes Veramo the chosen technology, as it allows the implementation of the Edge Agent with the independence of the chosen blockchain platform. Finally, we select React Native programming language, as this language allows the implementation of native applications (iOS and Android) to be deployed on personal devices. Veramo is implemented using JavaScript, and React Native is written in JavaScript, so we believe it is the best alternative for developing the user wallet, since this wallet will be deployed on each patient's and doctor's mobile phones.

\subsubsection{Cloud Agent}
This component enables communication between edge agents, interactions, and data retrieval from the blockchain. The Veramo framework also allows the implementation of a Mediator component to implement this functionality. The Mediator provides communication and interactions between decentralized agents or entities, acting as an intermediary that exchanges messages and data securely and efficiently. This covers the role of our cloud agent in generating peer-to-peer communications. To interact with the blockchain, Hyperledger Fabric implements a Client SDK to read and write data, and also to implement and deploy Smart Contracts (Chaincodes) on the blockchain.

\subsubsection{Mutual authentication and peer-to-peer communication}
Decentralized Identifier Communication (DIDComm) is a communication protocol that enables secure, private messaging between parties identified by DIDs. This is the protocol chosen to implement mutual authentication and peer-to-peer communication. On the one hand, mutual authentication is envisioned to be produced in a physical medical consultation. Here, the patient and doctor generate a DID, and one shows a QR code for the other to read. The messages are then sent through the Mediator from one party to the other, and the connection is established. This process creates a secure channel between these two entities thanks to the DIDs, the keys associated with such DIDs, and the DIDComm protocol. Veramo implements the DIDComm protocol, enabling the delivery of this communication between user wallets.

\subsubsection{Health data storage}
For the PoC, we execute the exchange of health data as VCs. To store them, Veramo implements TypeORM \cite{TypeORM} framework. TypeORM implements an Object Relational Mapping (ORM) framework. In essence, the utility of ORM is the mapping between entities and database tables, abstracting us from the relationship between tables. In the background, Veramo deploys TypeORM in conjunction with SQLite, a relational database. In terms of security, the database is encrypted with a strong password. Thanks to this storage, we can store either VCs, for the health data, and pairwise DIDs created for the different connections between patients, practitioners and laboratories.

\subsection{PoC Implementation}
\label{Testing}
With the core components defined in terms of specifications, we present the implementation performed on the PoC created. The code of this PoC is available in a public GitHub repository \cite{GitHubRepository}. To begin with, Table \ref{tab:technologies} shows the technologies and versions used for the PoC. The executions were performed with two edge agents, a cloud agent (mediator), and the blockchain platform. The edge agents were one physical smartphone (Samsung Galaxy M33 with 6GB RAM and Octa-core processor) and an emulated device (Pixel 4a with 6GB RAM and Snapdragon 730 processor), both with Android version 14. The cloud agent was a server implementation of the Veramo library, in collaboration with Fabric SDK, to manage the communication between the edge agents and store and retrieve data from the ledger. Finally, the blockchain platform, as commented on Section \ref{blockchain-platform}, was Hyperledger Fabric, using the test execution that it provides in its official repository. We also provided a Smart Contract that allows the storage and retrieval of DIDs.

\begin{table}
    \centering
    \caption{Technologies used in the PoC.}
    \begin{tabular}{|*{3}{c|}}
    \hline
    \textbf{Component} & \textbf{Technology} & \textbf{Version} \\ \hline
    Blockchain platform & Hyperledger Fabric & 2.10.0 \\ \hline
    Blockchain API & Fabric SDK & 1.4.0 \\ \hline
    Edge Agent & Veramo Framework & 5.6.1 \\ \hline
    Cloud Agent (mediator) & Veramo Framework & 5.6.1 \\ \hline
    Programming Language & React Native & 0.73.4 \\ \hline
    App test environment & Expo Go & 2.30.11 \\ \hline
    App operating system & Android & 14 \\ \hline
    Physical server & MacBook Pro & 14.4.1 \\ \hline
    \end{tabular}
    \label{tab:technologies}
\end{table}

\subsubsection{PoC workflow}
\begin{figure}[ht]
    \centering
    \includegraphics[trim = 0mm 280mm 0mm 40mm, clip, width=4.2cm]{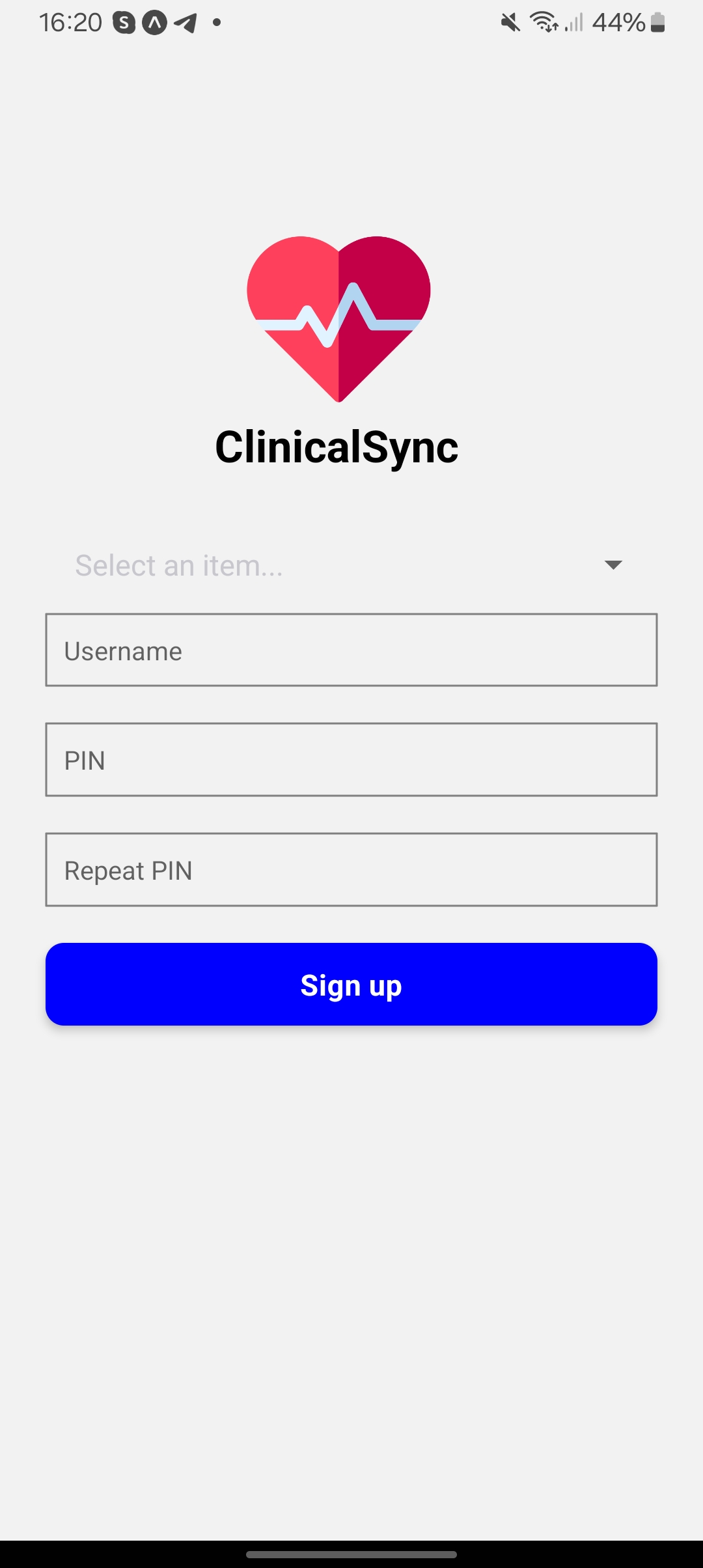}
    \includegraphics[trim = 0mm 280mm 0mm 40mm, clip, width=4.2cm]{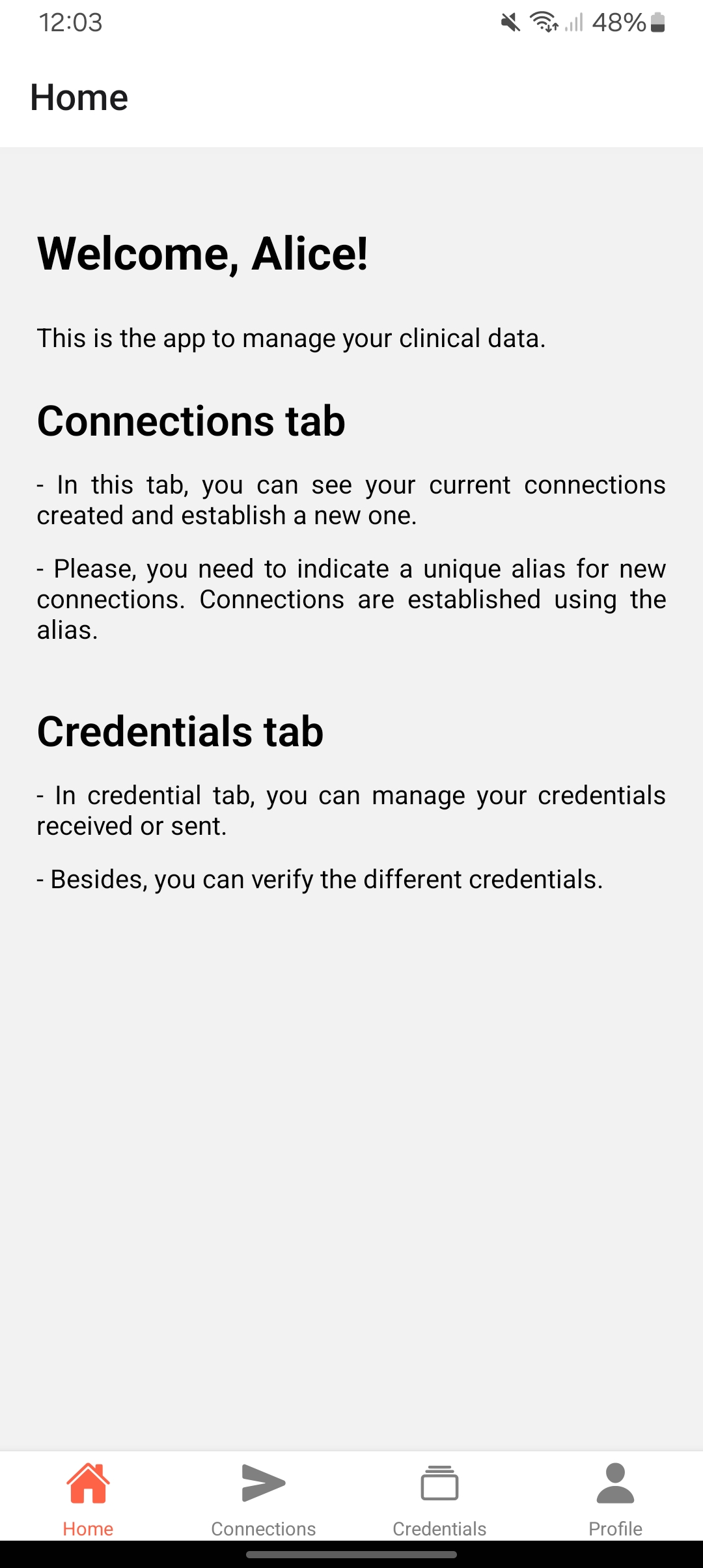}
    \caption{Registration and Home screens.} 
    \label{fig:register-home}
\end{figure}

In the PoC, we had two actors, a patient Alice and a doctor Bob. They represented the two edge agents. We represent a medical appointment between Alice and Bob, where they establish a secure connection, and Bob provides a prescription for Alice with a VC. The first step is to register these users. In the PoC, registration creates a username and PIN code to log in against the cloud agent (since we share the same cloud agent for Alice and Bob) and an X.509 certificate to interact with the Blockchain platform. Figure \ref{fig:register-home} shows the registration process and home screens. In the background, when Alice registers, her application creates a DID (anywise DID), sends it to the cloud agent, and the cloud agent connects to the SC installed in the blockchain to store her DID for future verification, as we explain below. In the same manner, we performed this step with Bob, and, at this point, we had two registered users in the framework. One important aspect regarding the DIDs is the DID method used in the PoC, that is, ``did:peer:2'' \cite{PeerMethod}. The reason to use this method is that it allows us to include the cloud agent endpoint directly, ensuring that the two edge agents can communicate with each other's cloud agent to verify the DID.

\begin{figure}[ht]
    \centering
    \includegraphics[trim = 0mm 15mm 0mm 40mm, clip, width=4.2cm]{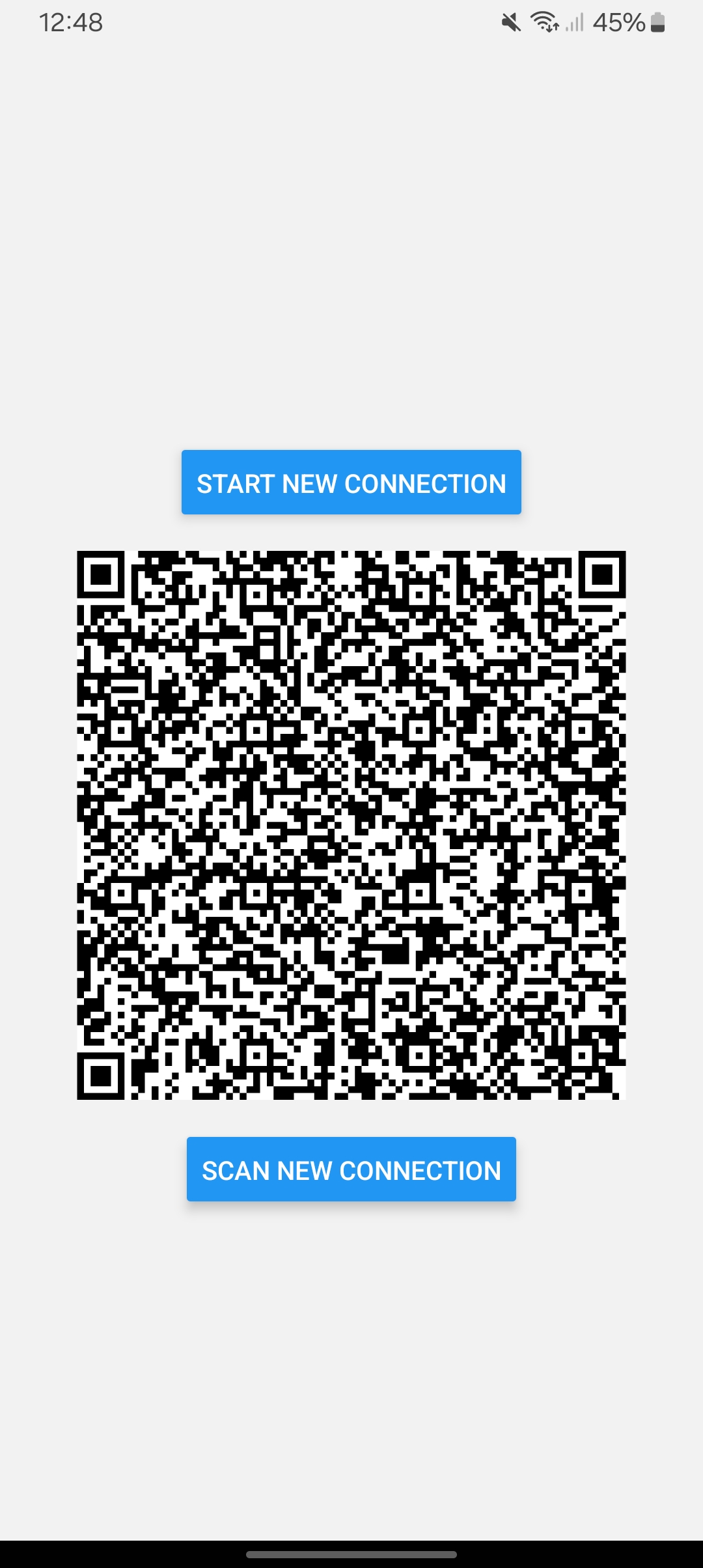}
    \includegraphics[trim = 0mm 15mm 0mm 40mm, clip, width=4.2cm]{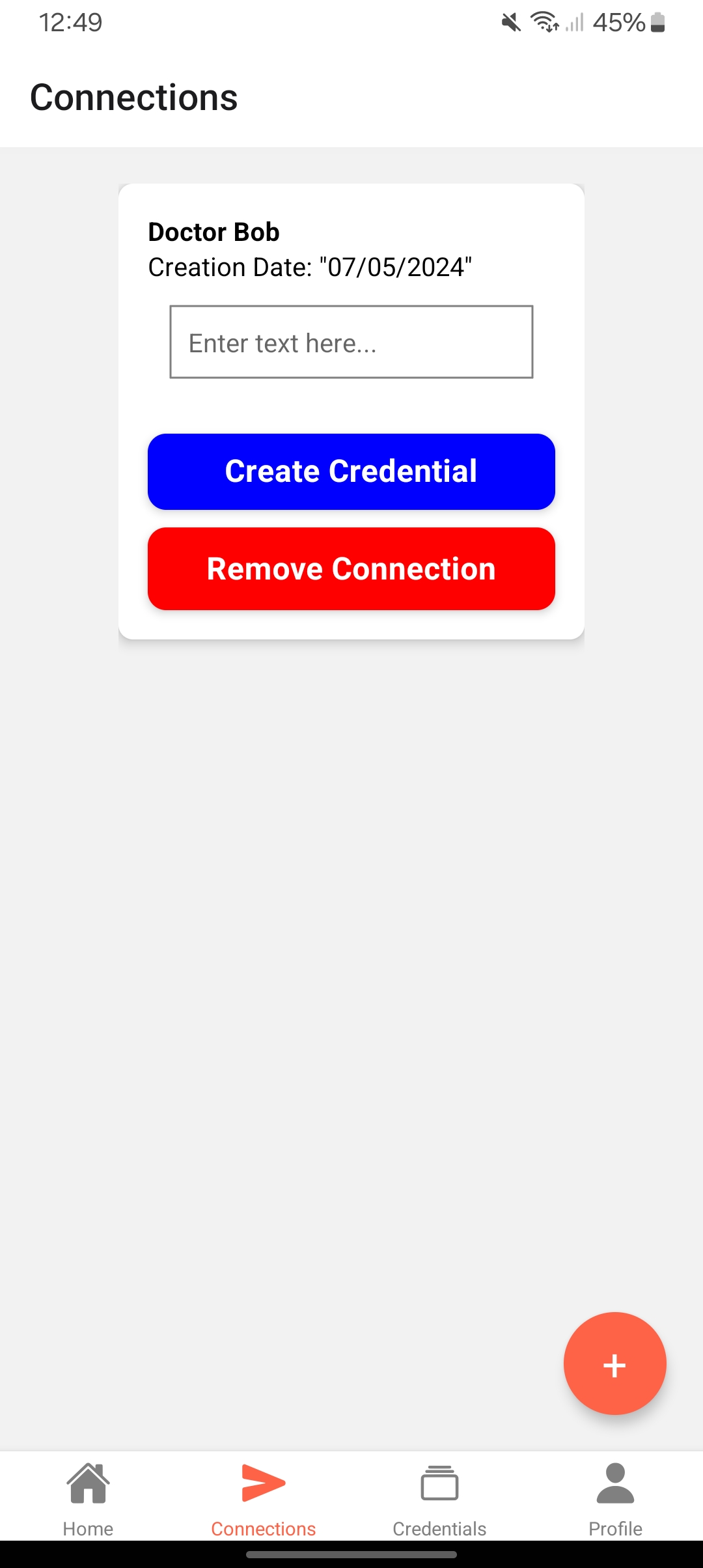}
    \caption{New connection and Connections screens.} 
    \label{fig:connections}
\end{figure}

Now, we are going to connect between Alice to Bob. To do this, we go to the Connections screen. There is a button for adding a new connection. Two steps are required to create a new connection. First, the application creates a DID (pairwise DID) with the alias that the user specifies for the connection. For example, Alice names the connection ``Doctor Bob'' and Bob ``Patient Alice''. The alias must be unique because it is associated with the DID created. These DIDs are known  only to Alice and Bob. Second, one user must create a QR and the other must scan a such QR. This triggers the connection to the cloud agent and establishes the connection between them. These screens are shown in Figure \ref{fig:connections}. There is an important aspect here: when a DID is created, the edge agent (as a recipient actor) connects to the cloud agent (as a mediating agent), indicating that it needs the cloud agent to act as a mediator for such DID. This is a protocol called ``Coordinate Mediation'' \cite{CoordinateMediation}, which is included in the DIDComm protocol. With this mediation process created, we need another protocol to retrieve the messages from the mediator to the recipients (Alice and Bob). The ``Message Pickup'' protocol \cite{MessagePickup} supports this process. This protocol consists of the cloud agent storing the messages waiting for the recipient to pick them up. Therefore, in the background, the process of establishing a new connection involves creating two pairwise DIDs, connecting to the cloud agent using the coordinate mediation protocol, and sending an initial message from  the QR code reader; for instance, Alice indicating that she scanned the connection and sent her pairwise DID created to Bob, who picks up the message from the cloud agent.

\begin{figure}[ht]
    \centering
    \includegraphics[trim = 0mm 240mm 0mm 35mm, clip, width=4.8cm]{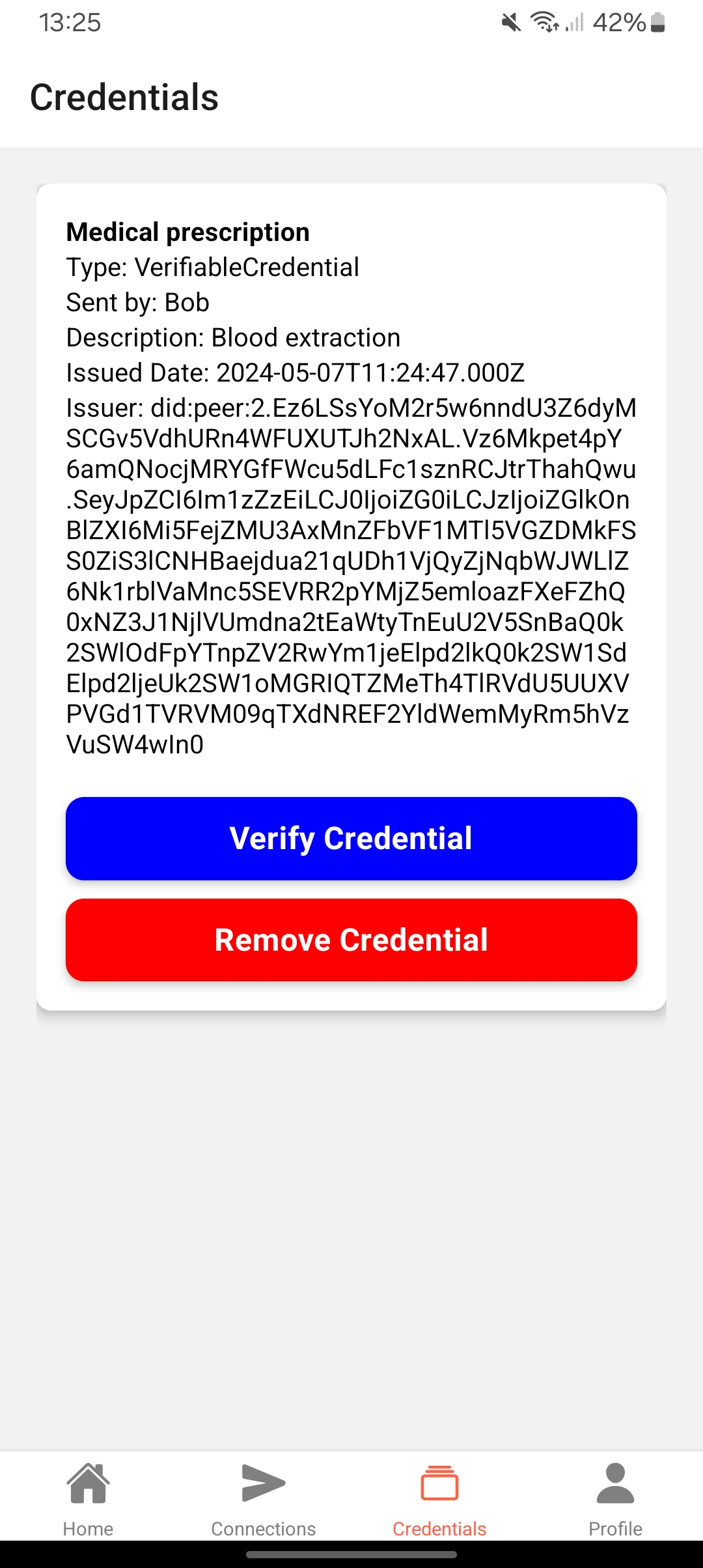}
    \includegraphics[trim = 0mm 240mm 0mm 35mm, clip, width=4.8cm]{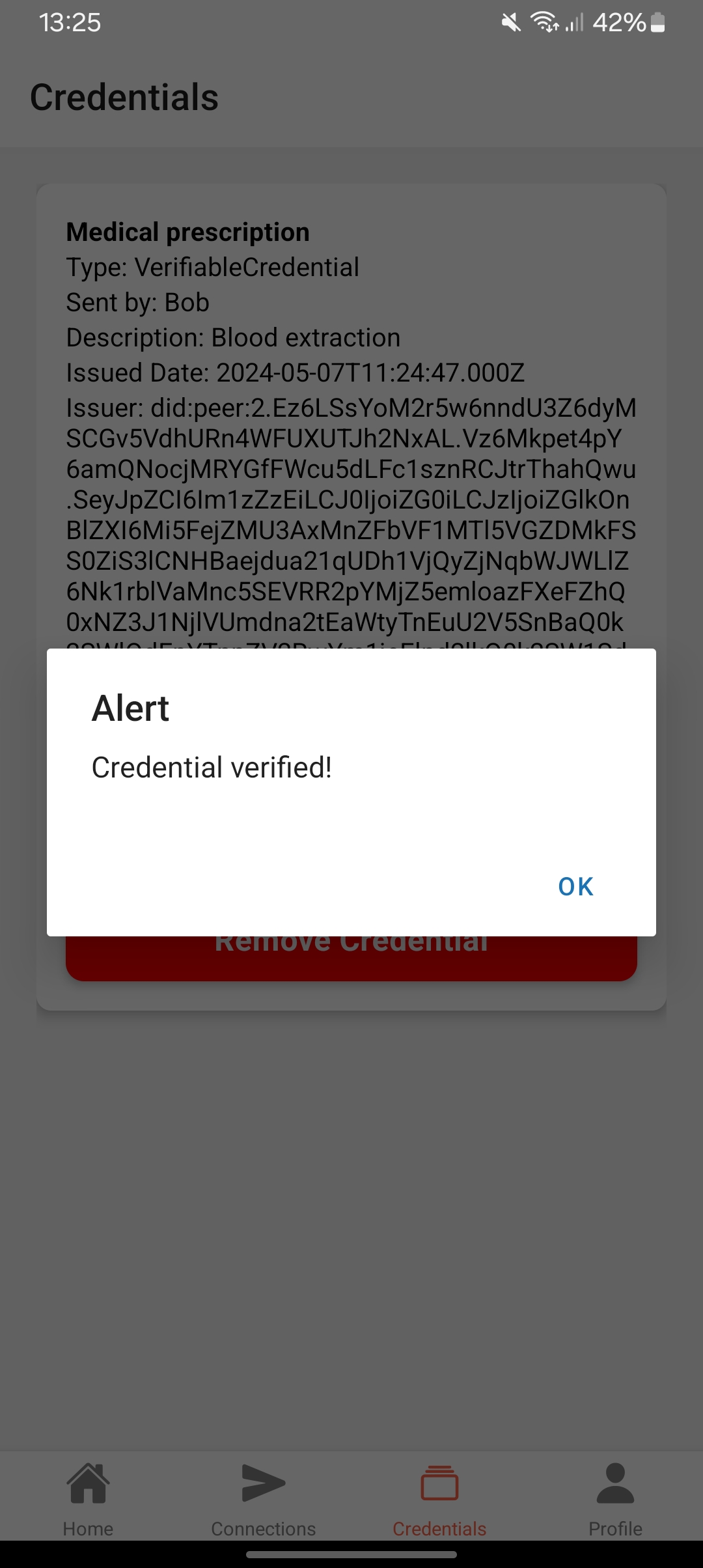}
    \caption{Credentials screen and verification process.} 
    \label{fig:credentials}
\end{figure}

With this mutually authenticated channel, both edge agents can send VCs. A VC must be signed by a DID, and this is where the DID stored in the blockchain during the registration process becomes relevant. Thinking in terms of a prescription, if Bob provides a new VC for Alice indicating that she needs a blood draw, the nurse in charge of such a process should be able to verify that this information is real. However, if we use the pairwise DID created for the connection, the nurse would not be able to verify Bob's identity. Therefore, each user creating a VC in the framework uses the anywise DID created in the registration process, which uniquely identifies them. With the signed VC, the nurse edge agent can go to the blockchain and verify that this DID exists and belongs to a valid user in the framework. Figure \ref{fig:credentials} shows the credentials screen with a VC received from Bob and the verification check performed with such VC.

These are all the design functionalities that we have implemented and tested. While the PoC  covers only certain aspects of the comprehensive design framework, it provides a robust foundation and validation of the overarching concept. The successful implementation of these initial components demonstrates the feasibility and potential of our approach. This PoC marks the critical starting point from which we can build, leveraging its successes to guide the development of the complete design framework.


\section{Performance tests}
\label{sec:tests}
We execute different performance tests to evaluate the functionalities and features provided in the PoC, and thus the feasibility of the framework. These tests focus on extracting the time required to create a DID/VC, establish a connection between edge agents, send DIDComm messages (establish mutually authenticated connections, exchange VCs, etc.), verify a VC, and write/read the DID available in the blockchain platform.

First, we distinguish between daily and infrequent operations to assess whether the cost of the operations is acceptable or not. Frequent/daily operations include: ``Create a VC'', ``Verify a VC'', ``Exchange DIDComm Messages'', and ``Read DID (Blockchain)''. Rare operations include: ``Create a DID'', ``Establish Connection'', ``Write DID (Blockchain)''. 

Second, for each operation, we measure the average time (in milliseconds) for 10.000 operations. Note that we are mainly working with edge agents running on mobile devices. Therefore, we need to ensure that the user experience is not affected by the different operations.

For the Blockchain operations (Read and Write DIDs), we use Hyperledger Caliper \cite{Caliper}, a benchmark tool provided by Hyperledger for testing chaincodes deployed on Hyperledger Fabric. This tool allows the execution of chaincode transactions to evaluate its performance. The blockchain test also considers a total number of 10,000 transactions and an issue rate of 200 transactions per second (TPS). Knowing that the writing of a DID is executed once on the first identity record (DID anywise), we are testing with 200 user records per second, which represents a realistic situation.

\begin{figure}[ht]
    \centerline{\includegraphics[width=0.95\textwidth]{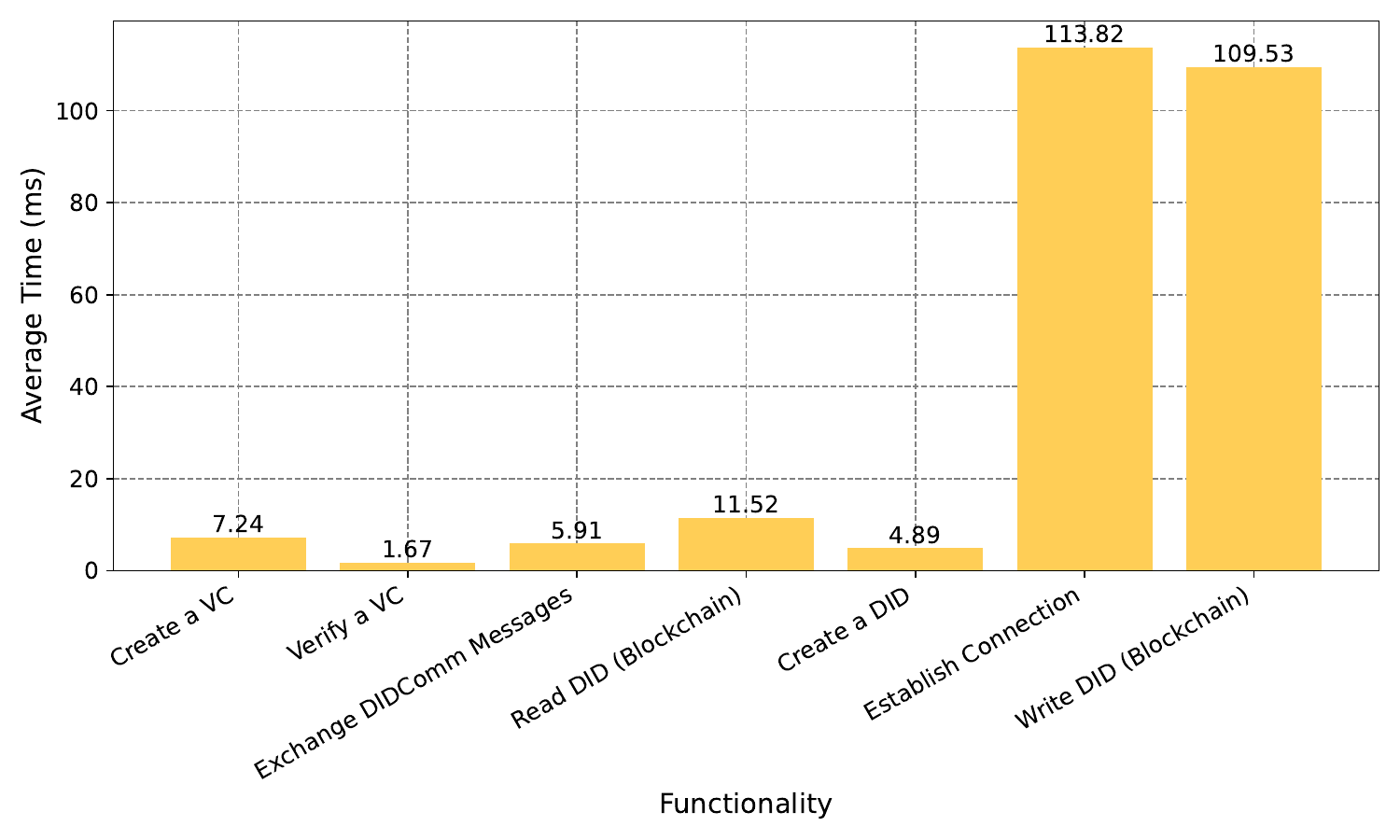}}
    \caption{Performance tests obtained for the SSI functionalities.} 
    \label{fig:performance-tests}
\end{figure}

Figure \ref{fig:performance-tests} shows the results obtained for the different operations. On the one hand, creating DIDs/VCs, verifying VCs and exchanging DIDComm messages take less than 10ms, a great result for realistic environments. For the case of establishing a connection, the results are 100ms due to the double DIDComm message required and the need to update the internal mediator table to add the new DID received for the ``Coordinate Mediation'' procedure, as commented in the previous subsection. This is not an issue because 100 milliseconds is not noticeable to the user, and the connection is made only at the first medical appointment between a patient and a healthcare professional; from that point on, the connection is established, and they can exchange messages over it. In addition, we can improve the results by studying the optimal number of edge agents per cloud agent, since not all platform users will use the same cloud agent. On the other hand, the tests performed with the chaincode implemented for storing and retrieving DIDs require 10ms for reading tasks and 100ms for writing tasks. In the case of writing DIDs, this is not an issue as it only happens once when the user registers because the pairwise DIDs are not registered yet in the blockchain platform. 

With these results, the PoC shows excellent performance in operations and transactions, indicating that the framework could be used with a high number of users. Thanks to this evaluation, we can conclude that the different healthcare actors and patients can interact between them with a good user experience, thanks to the reasonable time required for each operation and transaction.


\section{Discussion}
\label{Discussion}
Our solution extends the literature presented in Table \ref{tab:relatedwork} on three main pillars, as follows:

\begin{itemize}
    \item Data storage: There are different aspects to consider in a SSI solution. First, patients must manage and consent to access their data. This fact puts the patient at the center of the solution. However, where their data is stored is another aspect that many research efforts do not consider. Patients can control access to their data, but if they do not control storage, the data could be accessed without their consent. For example, malicious insiders at the data custodian could misappropriate the clinical information. Therefore, our solution also proposes patients as the custodian of their data by locating them in their personal devices. In this context, only the work of  Harrel et al. \cite{Harrell2022} proposed a framework where the patients store their own information.
    \item Recovery and emergency plan: The location of health data supports the fundamentals of SSI, but we may have a problem if patients lose access to their personal devices. For this, our solution includes a complete backup mechanism, that can be triggered with a social recovery protocol presented in Section \ref{wallet-recovery}. We also consider an emergency situation where the patient cannot consent to access the data. In this case, we also propose to use such a social recovery protocol to give access to the doctor and to record the related  log with the accessed data in the blockchain ledger, to avoid malicious activities with the patients' data. From the literature review, we concluded that only two works considered an emergency situation \cite{Saidi2022, Tcholakian2023}, and only one defined a backup mechanism for recovering health data \cite{George2023}.
    \item Blockchain platform: The literature explores the use of Hyperledger Indy for implementing SSI solutions, as this blockchain is optimized for managing identity information (DIDs, VCs, etc.). However, our framework takes a different approach by using Hyperledger Fabric. This platform is not focused on SSI, but it provides a number of characteristics that Indy does not. On the one hand, Fabric allows the implementation of smart contracts, providing flexibility, scalability, modularity, and the potential to create any logic requested. On the other hand, Fabric offers the provisioning of storage channels that are accessed based on the role received in the platform. This implies the ability to design any use case and workflow, from storing DIDs to sharing data with third parties for research purposes. In reviewing the literature, only the work of Tcholakian et al. \cite{Tcholakian2023} uses Hyperledger Fabric to implement a SSI solution.
\end{itemize}

Therefore, our work provides a novel framework for protecting and securing access to health data, empowering patients to be the true owners of their own data. In this context, it is also imprative to highlight some limitations of the proposed framework. For example, implementing such a framework is challenging because the current health system is based on a centralized architecture. First, we should be able to export the patients' data to their personal devices. Second, we should change the infrastructure currently installed by the blockchain, deploying the blockchain nodes in strategic geographical zones, and establishing the different organizations around the regional/national/international field. Third, we should train and raise awareness among patients and healthcare/clinical professionals to understand how this solution works and its benefits \cite{Fernandez2024}. 

Fourth, we should provide the framework to real users to analyze and study user acceptance. Fifth, we envision a privacy issue for users with limited knowledge about privacy, since healthcare professionals may ask for more data than they need. Moreover, we could have a significant security issue if the CA or MSP acts maliciously, as this would compromise the whole framework. However, our proposal represents a starting point for providing and implementing decentralized and secure technologies that eliminate the single point of failure and distribute the resources and data among all stakeholders involved in the framework.


\section{Conclusion}
\label{Conclusion}
Health data is one of the most sensitive data in society. Therefore, such information should be protected and secured with innovative and effective technologies. In this article, we have reviewed the related work on SSI frameworks for healthcare and clinical environments. We found that various solutions have been presented, but a complete definition that includes innovative and relevant functions for this use case is still needed. To address this challenge, we created a novel open-source framework by providing a secure edge-cloud agent and a specific blockchain-based infrastructure that empowers patients as owners of their data, by storing and managing it on their personal devices. This framework enables self-management of the healthcare domain, creating secure connections and authenticated channels between patients and healthcare professionals, thanks to pairwise DIDs. Moreover, verifiable information can be exchanged using VCs and VPs, ensuring the patient-centric approach promised.

Furthermore, we have validated the feasibility of our framework with a proof-of-concept (PoC) simulating a real interaction between a patient and a doctor, creating a secure connection thanks to an authenticated channel between them, storing and retrieving data from the blockchain platform, and securely exchanging VCs. This procedure satisfies the SSI functionality defined in the framework. Moreover, we envision it evolving into an open-source solution. This approach not only promises to inspire future projects but also aims to continually enhance and refine the current framework. 

As perspectives for future work, we plan to extend our PoC implementation (e.g., framework functionalities and number of users), as well as the inclusion of privacy agent technology in the edge agent. A privacy agent assists the owner of the data in making appropriate decisions with respect to privacy. Patients want to manage their data privately, but often they are not experts or do not know how to apply effective privacy policies. To do this, the edge agent could act as a recommender to share only the relevant information based on the privacy policies implemented on it. In addition, the edge agent could act as a rewriter of queries, taking a patient query as input, redefining it to be privacy-friendly and applying efficient selective reading. This could significantly advance the proposed framework and initiate another line of research.


\medskip

\noindent \textbf{Acknowledgments ---} This work has been partially funded by the strategic project CDL-TALEN TUM from the Spanish National Institute of Cybersecurity (INCIBE) by the Recovery, Transformation and Resilience Plan, Next Generation EU. Furthermore, this work benefited from State aid managed by the Agence Nationale de la Recherche under the France 2030 programme, reference ANR-22-PESN-0006, project TRACIA. The work has also been partly supported by the chair Values and Policies of Personal Information, Institut Mines-Telecom, France; and by the International Alliance for Strengthening Cybersecurity and Privacy in Healthcare (CybAlliance, Project no. 337316).

\bibliographystyle{elsarticle-num}
\bibliography{bibliography}

\end{document}